\documentclass[10pt,a4paper]{article}
\usepackage[utf8]{inputenc}
\usepackage{amsmath}
\usepackage{amsfonts}
\usepackage{amssymb}
\usepackage{graphicx}
\usepackage{fullpage}
\usepackage{booktabs}
\usepackage{float}
\usepackage{caption}

\title{A Survey of Automated Programming Hint Generation - \\ The HINTS Framework}
\author{Jessica McBroom, Irena Koprinska\thanks{equal contribution} \ and Kalina Yacef\footnotemark[1] \\  \texttt{\{jmcb6755, irena.koprinska, kalina.yacef\}@sydney.edu.au} \\ \\
School of Computer Science, University of Sydney }
\date{}

\begin{document}
	\maketitle
	
	\begin{abstract}
		Automated tutoring systems offer the flexibility and scalability necessary to facilitate the provision of high quality and universally accessible programming education. In order to realise the full potential of these systems, recent work has proposed a diverse range of techniques for automatically generating hints to assist students with programming exercises. This paper integrates these apparently disparate approaches into a coherent whole. Specifically, it emphasises that all hint techniques can be understood as a series of simpler components with similar properties. Using this insight, it presents a simple framework for describing such techniques, the \underline{H}int \underline{I}teration by \underline{N}arrow-down and \underline{T}ransformation \underline{S}teps (HINTS) framework, and it surveys recent work in the context of this framework. It discusses important implications of the survey and framework, including the need to further develop evaluation methods and the importance of considering hint technique components when designing, communicating and evaluating hint systems. Ultimately, this paper is designed to facilitate future opportunities for the development, extension and comparison of automated programming hint techniques in order to maximise their educational potential.
	\end{abstract}

\section{Introduction}
Automated tutoring systems, which provide educational materials and feedback to students without direct teacher involvement, offer promising approaches to delivering scalable and high-quality programming education. One fundamental aspect of these systems is the provision of hints and guidance to students working on programming tasks. Specifically, automated hints can help students progress in their learning by providing instant and relevant feedback to correct their mistakes and help them advance through exercises. In recent years, numerous techniques for producing programming hints have been developed, including approaches aimed at scaling up instructor feedback \cite{Head-GSSFDH:2017,Lee-YTWP:2018,Haldeman-TBBSYN:2018}, extracting patterns from peer data \cite{Lazar-MB:2017,Iii-HB:2014,Lazar-SB:2017}, identifying particular dynamic or static issues with student programs \cite{Edmison-E:2015}, automatically generating personalised paths to solutions \cite{Keuning-HJ:2014,Gerdes-HJV:2017,Price-ZB:2017,Rivers-K:2017}, or combinations of these \cite{Piech-SHG:2015, Price-DB:2016}. 

While this diversity of approaches offers a great range of potential options for improving feedback, it also presents a difficult challenge - namely, it is difficult for instructors and researchers to decide which techniques are most effective for different situations. To address this issue, many studies have employed a range of different evaluation methods, including user studies and surveys \cite{Antonucci-ENPM:2015, Gerdes-HJV:2017, Lazar-SB:2017,Head-GSSFDH:2017,Lee-YTWP:2018}, comparisons with experts \cite{Piech-SHG:2015, Price-ZB:2017}, analysis using historical data \cite{Keuning-HJ:2014, Iii-HB:2014, Edmison-E:2015, Price-DB:2016, Gerdes-HJV:2017, Lazar-MB:2017, Marin-PSR:2017,Rivers-K:2017,Chow-YKC:2017,Haldeman-TBBSYN:2018} or other technical evaluations \cite{Marin-PSR:2017}. Researchers have also conducted comparison studies on small subsets of techniques \cite{Price-DZPLCB:2019,Piech-SHG:2015}. However, the time-consuming nature of evaluations, complexity of hint techniques and the apparently disparate nature of hints produced by these techniques (e.g. for different purposes, programming languages and students) precludes a comprehensive study of every technique to achieve unambiguous answers. In addition, this difficulty is exacerbated as more hint techniques and variations are developed. Overcoming these challenges requires a clearer theoretical perspective to draw these techniques together into a simple, coherent picture. Such a perspective could motivate more focused questions for empirical studies to investigate, facilitate the sharing of ideas across hint techniques and provide teachers and researchers with insight into the range of available approaches and their relationships with each other. As such, it would act as an important step towards discovering the most useful techniques for different situations, in order to maximise the effectiveness of programming hints.

As a step towards addressing these challenges, this paper surveys and develops key theoretical ideas behind recent work from the last five years (2014-2018) on generating automated hints for programming exercises. Specifically, it draws together hint techniques into a common framework that is highly generic and modular, yet simple. In addition, it discusses recurring themes and investigates how these can inform future work. As such, it acts as a guide for understanding the important developments, challenges and future directions in the field of automated programming hints, with the ultimate goal of maximising the potential of automated tutoring systems.

This paper is set out as follows. Section \ref{sec: related_work} discusses related reviews and surveys, with an emphasis on how this work contributes to the field. Section \ref{sec: scope} defines the scope of this survey. Section \ref{sec: HINTS Framework big} presents the \underline{H}int \underline{I}teration by \underline{N}arrow-down and \underline{T}ransformation \underline{S}teps (HINTS) framework for understanding hint generation techniques, then Section \ref{sec: example guided review} surveys such techniques in the context of this framework. Finally, Section \ref{sec: insights} concludes with a discussion of key insights and future directions in the field.

\section{Related Surveys and Reviews} \label{sec: related_work}

Previous work on categorising automated programming feedback has generally used one of three criteria to distinguish between feedback classes:
\begin{enumerate}
	\item the \textbf{technique} used (i.e. how the feedback is produced). For example, Markov Decision Processes (MDPs), Bug libraries or test cases could be considered different techniques for producing hints.
	\item  the \textbf{nature} of the feedback (i.e. the type of information it reveals to students). For example, the feedback may be directed towards different parts of the program (syntax, layout, output), may be very specific or general, or may be targeted in different ways (e.g. towards mistakes or towards next steps to try).
	\item the \textbf{input} required (i.e. what data is used to produce the feedback). For example, the feedback may be produced using model solutions, peer data or test cases as input, and the format of this data could be different too. For instance, programs could be input as abstract syntax trees (ASTs) or lines of code.
\end{enumerate}
In \cite{Le:2016}, a survey of adaptive programming feedback, feedback is categorised based on its nature. Specifically, it is divided into the classes ``yes/no", ``syntax", ``semantic", ``layout" and ``quality", where``yes/no" feedback reveals whether work is correct, and the other classes reveal information about syntactic, semantic, layout or quality (e.g. efficiency) issues respectively. In contrast, in \cite{Striewe-G:2014}, which reviews static analysis approaches to producing feedback for Java programs,  tools are classified based on the input used. For example, tools are classified based on the number of files they accept (e.g. ``Single File vs. Multi File Analysis") and the program representations they require as input (e.g. ``Trees vs. Graphs", "Source Code vs. Byte Code Analysis").

In \cite{Keuning-JH:2018} (extended from \cite{Keuning-JH:2016}), all three types are used separately to label  programming tools. The tools are labelled based on the technique used (e.g. ``Model Tracing", ``Data Analysis", ``Program Transformations"), the nature of the feedback (e.g. ``Knowledge about Task Constraints", ``Knowledge about Concepts", ``Knowledge about Mistakes") and also separately based on their ``adaptability" (i.e. the input required by the system, such as ``Solution Templates", ``Model Solutions" or ``Test Data").

Since the purpose of these previous surveys has generally been to compare feedback tools, they focused on dividing feedback into general categories so these tools could be distinguished. In contrast, while our work also uses the technique to classify feedback and also makes reference to the input and nature of the feedback, our focus is not on assigning feedback techniques to categories, but instead on building one simple and integrated picture of these. Our work considers the individual components that comprise these techniques, and utilises these to draw out insights about the nature of hint generation. As such, it conducts a deeper exploration into the nature of programming hints.

Our work is situated more broadly in the area of automated programming tutors. Other reviews in this area have considered techniques for automated assessment (e.g. \cite{Ala-K:2005} and \cite{Ihantola-AKS:2010}), or approaches to tutoring that can be supported by AI techniques \cite{Le-SGP:2013}, such as ``example", ``simulation" or ``dialogue"-based approaches. In addition, some work has focused on the general features of programming tutors, such as the programming languages taught, or primary and supplementary features \cite{Crow-LW:2018}. While some of these reviews also reference automated programming feedback, it is generally brief or not the main focus.

Our work also relates to reviews on automated software repair and debugging. For example, \cite{Monperrus:2018} presents a bibliography of automated software repair techniques and \cite{Silva:2011} surveys algorithmic debugging strategies. While our work also reviews some techniques for debugging and repairing programs, the focus is on producing hints for students in an educational context, where different resources may be available to the system, such as data from peers or teachers.

\section{Scope} \label{sec: scope}
The scope of this survey is recent (2014-2018) techniques for generating automated hints for programming exercises:

\begin{enumerate}
	\item We consider a \textit{programming hint} to be any type of feedback that improves a student's knowledge of how to complete a programming exercise. For example, it may help them to identify mistakes in their program, suggest potential ways to proceed, recommend concepts to revise or clarify the task requirements. However, it does not, for example, include feedback aimed at encouragement or emotional support. 
	\item We define \textit{automated hints} very broadly to be any hints where no human intervention is required between the time the hint is requested and the time the hint is given. As such, it can still include hints generated using historical peer data, pre-written teacher hints, or other resources produced by people, so long as human intervention is not required when the actual hint is produced. However, it does not include peer-to-peer hints or teacher hints written \textit{after} the hint is requested.
\end{enumerate}

Note that this paper focuses on methods for \textit{generating} hints. For this reason, work on hint timing (e.g. based on student emotion \cite{Tiam-S:2018, Barron-ZHB:2015}), restricting hint availability (e.g. to prevent hint overuse \cite{Annamaa-SV:2017}) or deciding which hint technique to use for a particular student (e.g. using student models) are beyond the scope of this paper.

Note also that, while the focus of this paper is automated hints and not automated grading, sometimes papers on automated grading are discussed if the grading technique could also be used to produce hints. For example, test cases can be used to grade student programs, but can also be used to give hints about the types of inputs the program is failing on.

\section{The HINTS Framework for Generating Programming Hints} \label{sec: HINTS Framework big}
A diverse range of approaches to generating automated programming hints have been proposed in recent years. These approaches are based on a variety of ideas and techniques, including machine learning, the utilisation of peer or teacher data, debugging techniques and other methods focused on diagnosing errors or discovering potential improvements in student programs. In addition, these techniques result in hints in various forms, including hints that highlight errors, direct student actions, recommend additional materials or provide other forms of support.

In order to understand how this multitude and variety of hint techniques fit together, existing approaches based on general categories are not sufficient, for the following reasons:

\begin{enumerate}
	\item categorisation approaches based on the general \textbf{technique} employed are problematic because techniques are so readily combined. For example, even if two approaches were fundamentally different, it would be possible to integrate the output from both, thereby producing a third approach that fit neither category. Indeed, difficulties with this type of categorisation have been reported in previous investigations such as \cite{Keuning-JH:2018}, where around 28\% of surveyed tools did not suit a particular technique category.
	\item categorisation approaches based on the \textbf{nature} of hints produced are also problematic, because hint type does not necessarily correspond to the technique used. For example, a hint to delete a line of code could be produced by identifying it in a library of common bugs, as a frequent action taken by peers or as a step towards a model solution - all different techniques. Conversely, even if a problematic line of code was identified in the same way, the final form of hints could be quite different: they could explicitly instruct the student to delete the line, recommend reading materials relevant to it or perhaps identify its rough vicinity in order to focus the student's attention. As such, vastly different hint techniques can produce hints of a similar nature, and similar hint techniques can produce hints of vastly different nature. For these reasons, the final form of hints is also not ideal for understanding hint generation techniques.
	\item categorisation approaches based on the \textbf{input} are also problematic because the same input can be processed in many different ways to achieve vastly different results. For example, data from peers could be used to find common paths to a solution or mined to find common buggy patterns.
\end{enumerate}

If these categories are not sufficient, this suggests that hint techniques may be too complex to allow for easy categorisation when considered in their entirety. However, hint techniques are often comprised of many smaller steps that are each simpler than the entire technique. This prompts the following question: 

\begin{center}
	\textbf{Can we develop a simple framework to describe all hint techniques by considering the smaller steps they are comprised of?}
\end{center}

A framework describing how automated hints are produced would be an important step towards understanding how these techniques relate and differ from each other, finding ways to extend and improve them, and developing  methodologies for evaluating and comparing them. As such, it would act as an important step towards realising the full potential of automated hint systems.

This section, and the following sections, argue that such a framework is indeed possible. Moreover, they demonstrate that recent hint approaches are built up from just two simple operations applied iteratively. Considering the diversity of techniques, this is both an important result and tool for understanding fundamental ideas behind automated programming hints.

Before presenting the framework, Section \ref{sec: themes in hint generation} introduces three examples of automated hint techniques and highlights key similarities in the processes that comprise them. This will act as motivation for the key ideas behind the HINTS (Hint Iteration by Narrow-down and Transformation Steps) framework for describing automated hint techniques in general, which will be presented in Section \ref{sec: framework for hint generation}.

\subsection{Themes of Hint Generation Techniques} \label{sec: themes in hint generation}

In order to explore several key ideas relevant to hint techniques, we first begin with a discussion of three examples of such techniques. Though these do not exemplify all approaches, they are illustrative of their diversity, and the similarities they exhibit will act as a basis for the general framework discussed in the next section. The examples are as follows:

\begin{enumerate}
	\item (MB) MistakeBrowser \cite{Head-GSSFDH:2017}. Using a database of program transformations learned from peer data, the system searches for a subset of these transformations that correct a student's program. If successful, it then presents the student with hints written in advance by teachers for this particular set (cluster) of transformations.
	\item (SFL) Spectrum based fault localisation  \cite{Edmison-E:2015}. Test cases are run against a student's program, comparing its output to the expected output on various inputs. After this, parts of the program (program spectra) that are used when the tests pass and fail are compared, and functions associated with failed tests are flagged and highlighted to the student as hints.
	\item (SC) SourceCheck \cite{Price-ZB:2017}. An incorrect student program is matched to the closest known solution based on a distance measure defined by the authors. The edits needed to convert the student's program to this solution are then presented as hints.
\end{enumerate}

While these examples all appear to be quite different, they all share an important similarity: they build up hints in \textit{levels} of increasing complexity. That is, they begin with simple hints, which are then developed into more sophisticated hints through an iterative process. For example, in (SFL) the final hint (highlighting a particular function) is produced using the results of test cases, which themselves could act as hints. Indeed, test cases are commonly used in programming tutors \cite{Keuning-JH:2018}. Moreover, these test case results are produced by comparing the output of the student's program to the expected output and, again, these themselves can act as simple hints revealing information about the student's program or the task respectively to the student. In this way, the hint technique can be considered a series of smaller steps, producing increasingly sophisticated hints. The levels of hints for all examples are shown in Table \ref{tab:hint_levels}.

\begin{table}[h!]
	\caption{Hint levels for each hint generation technique. The first column shows the input to the system. This is then processed to produce a new intermediate hint (second column). Finally, the intermediate hint is processed to produce another final hint (third column).}
	\label{tab:hint_levels}
	\begin{minipage}{\columnwidth}
		\begin{center}
			\begin{tabular}{lp{5cm}p{4cm}p{3cm}}
				\toprule
				\textbf{Example}		 &  \textbf{Input}	& \textbf{Intermediate Hint} 	& \textbf{Final Hint} \\
				\hline
				MB & All teacher hints, all transformations, student program, correctness test & Transformations that correct student program &  Most relevant teacher-written hint \\
				SFL & Student program outputs, expected outputs  & Passed/failed tests & Problematic functions\\
				SC & All solutions, student program  & Closest solution & Edits to solution \\
				\bottomrule
			\end{tabular}
		\end{center}
		\bigskip\centering		
	\end{minipage}
\end{table}

Surprisingly, not only can each of these examples be divided into simpler steps, but also these steps are remarkably similar: in each case, they involve deriving hints by \textit{narrowing down} a set of hints from an earlier level. For example, in (MB) the most appropriate teacher hint is derived by narrowing down the set of all teacher hints. In (SC), the closest solution is derived from the set of all solutions. In all cases, this \textit{narrowing down} operation also follows a similar pattern: hints are selected from the earlier set based on their relevance to the student's program and/or some quality criterion. For example, in (MB) the most appropriate transformations are selected from the set of all peer transformations based on whether (a) they can be applied to the student's program (relevance to the student's program) and (b) they are able to correct the program (quality). In (SFL) the important problematic functions are derived from the set of all functions based on their associations with passed and failed tests of the student's program (i.e. their relevance to the student program). In this way, the steps used to build up hints all follow a similar process. The steps for all examples are summarised in Table \ref{tab:step_examples}.

\begin{table}[h!]
	\caption{The steps used to derive hints in all three technique examples. Each step involves narrowing down a set of hints (hint data) from an earlier level. This is achieved by selecting hints from the set that (a) are relevant to the student's program and/or (b) satisfy a quality criterion.}
	\label{tab:step_examples}
	\begin{minipage}{\columnwidth}
		\begin{center}
			\begin{tabular}{llp{8cm}}
				\toprule
				\textbf{Example}	 & \textbf{Hint Data} & \textbf{Description}  \\
				\hline
				MB &  All peer transformations &  Select transformations that (1)  can be applied to the student program (\textit{relevance}) and (2) can correct the student program (\textit{quality}).
				\\
				\cline{2-3}
				& All teacher hints & Select teacher hints that are attached to the transformations that correct the student program (\textit{relevance}). \\ \hline
				SFL  & All student program output & (To determine incorrect output) select the output that differs from expected output (\textit{quality}) \\
				\cline{3-3}
				&  &  (To determine correct output) select output that  matches the expected output (\textit{quality}) \\
				\cline{2-3}
				& Student program & Select functions from the student program that  are mostly associated with incorrect output (\textit{quality}) \\
				\hline
				SC  & All solutions & Select solution that is closest to the student program (\textit{relevance})\\
				\cline{2-3}
				& Solution & (To discover code to add) select parts of solution that don't match the student program (\textit{relevance})\\
				\cline{2-3}
				& Student Program & (To discover code to delete\footnote{There are also hints to move code described in the paper, which can considered deletions followed by insertions.}) select parts of the student program that  don't match the solution (\textit{quality})\\
				\bottomrule
			\end{tabular}
		\end{center}
		\bigskip\centering		
	\end{minipage}
\end{table}

At this stage, the structural similarities between these three hint techniques may appear to be coincidental. However, this is not the case: they are, in fact, characteristic of recent hint techniques and can provide valuable insights into hint generation. These similarities will act as an essential basis for the framework describing automated hints presented in the next section.

Note that, in some of the examples shown in Table \ref{tab:step_examples}, the hint from an earlier level is not explicitly a set of hints, but is easily converted to one through simple transformations. For example, in (SFL) the student program is converted to a set of functions. In (SC) the solution and student program are divided into a set of smaller parts in order to find differences. In these cases, the transformations are trivial, but other techniques can involve more sophisticated transformations.

\subsection{The HINTS Framework for Hint Generation} \label{sec: framework for hint generation}
In the previous section, three examples of hint generation techniques were presented, with each able to be divided into a series of steps that produced hints in \textit{levels}. These steps involved \textit{narrowing down} and \textit{transforming} hints from earlier levels. Here we present the Hint Iteration by Narrow-down and Transformation Steps (HINTS) framework, which is based on these ideas.

\begin{figure}[h!]
	\centering
	\includegraphics[width=1\linewidth]{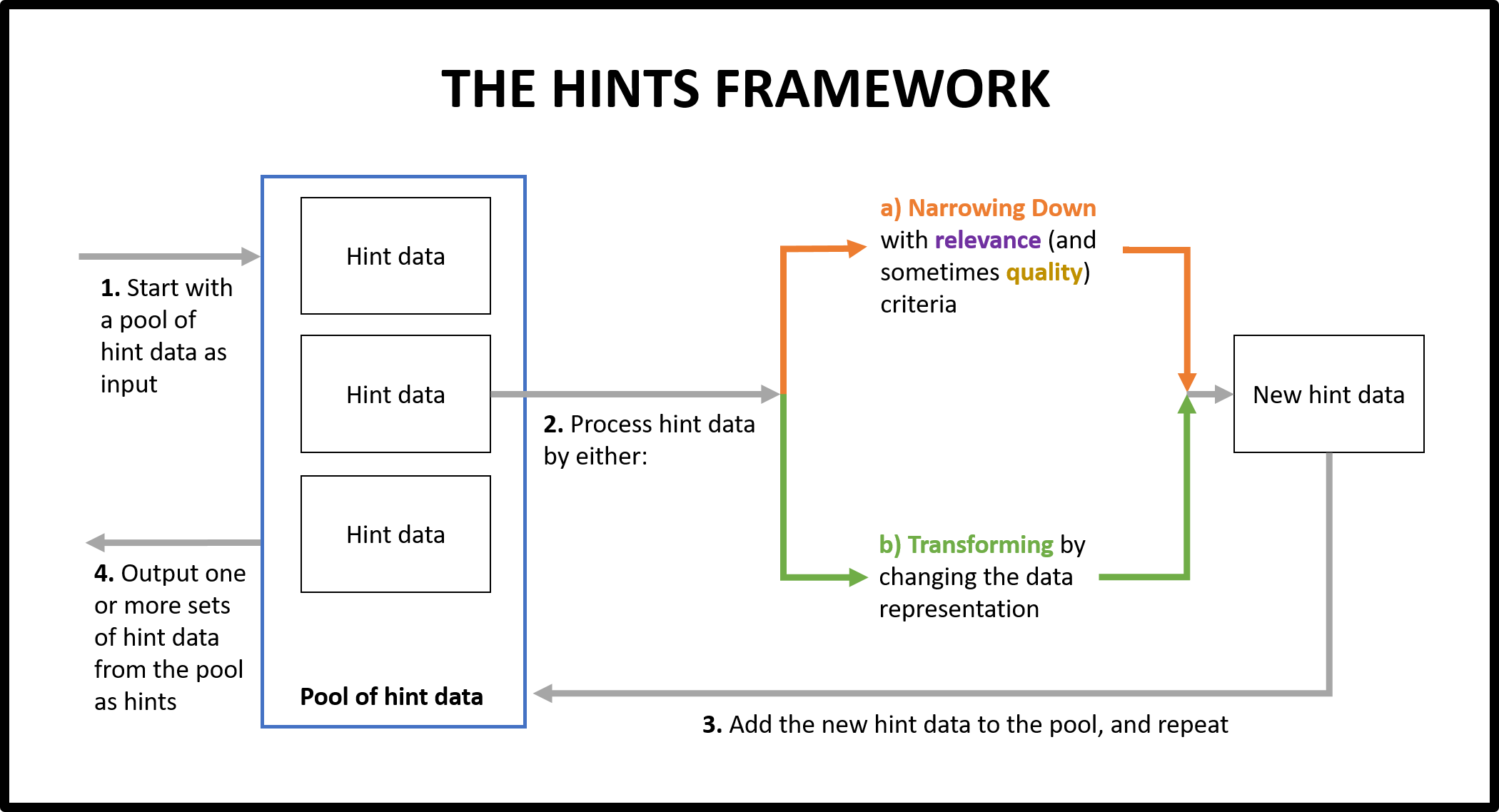}
	\caption{The HINTS framework for describing automated hint generation techniques. A small pool of hint data is given as input, and new hint data are produced by transforming or narrowing down existing hint data. Finally, some hint data is selected to be given as hints.}
	\label{fig:framework}
\end{figure}

The HINTS framework, shown diagrammatically in Figure \ref{fig:framework},  is described as follows:
\begin{enumerate}
	\item The system begins with a pool of data as input, which we call \textit{hint data}, since it is used to produce hints. Input hint data can include datasets produced by peers or teachers; features of the help-seeking student's program or submission history; or a correctness test, such as test cases, desired static program features or a model solution. While this input data is usually processed before being given to students as hints, note that it could potentially be used as a hint without further processing. For example, expected input/output from test cases could be given to students to help them understand the task requirements, and an entire dataset of teacher-written hints could be given to increase their awareness of common errors. 
	
	\item Hint data from the pool can be processed to produce new hint data. This is achieved by applying one of two steps:
	
	\begin{enumerate}
		\item a \textit{narrow-down} step. This involves taking a set of hint data and selecting a subset of this data based on a relevance criterion and/or a quality criterion. The relevance criterion stipulates that the subset of data should be relevant to some feature of the student's program. The quality criterion, in contrast, stipulates that hints should be of high enough quality (based on some measure). For example, in (MB) from Section \ref{sec: themes in hint generation}, the set of all peer transformations was narrowed down to a subset based on the following criteria: a) (relevance) that the transformations could be applied to (i.e. were relevant to) the student's program and b) (quality) that they were able to produce a correct solution. As another example, consider a system that takes a large set of peer program states and narrows these down to only the most popular reachable state from the current student's program. Then, the relevance criterion would be that the state is reachable from the current student's state and the quality criterion would be that the next state is popular.  
		
		\item a \textit{transformation} step, which involves changing the way hint data is represented. This could include dividing the hint data into parts, representing it with a different data structure or converting it to a standard form. For example, a model solution to a programming task could be divided into a set of functions, represented with a different data structure, such as an abstract syntax tree (AST) \cite{Lazar-MB:2017}, or converted to a canonical (standard) form through semantics-preserving transformations \cite{Gerdes-HJV:2017}. In addition, a set of student submissions could be divided into groups based on their output \cite{Iii-HB:2014}, or represented as a graph with nodes for programs and edges for transitions between them. 
	\end{enumerate}
	
	\item Once new hint data is produced, it can be added to the pool of available hint data, and the proceeding two steps can be applied iteratively to produce increasingly complex hints.
	
	\item One or more \textit{sets of hint data} from the final pool can be selected and offered to students as \textit{hints}.
\end{enumerate}

\section{An Example-Guided Survey of Hint Methods} \label{sec: example guided review}
In the previous section, the HINTS framework was presented to describe automated programming hint techniques. This section now reviews recent work in the context of this framework, showing how hint techniques fit into the framework, and also how they relate to each other.

In order to build a coherent overall picture of hint techniques, this review progresses through a series of stages, guided by example hint techniques from recent work. In each stage, an example that is different from anything discussed so far is presented and related to the framework. Then, other work extending upon or relevant to this example is introduced and discussed. Finally, all techniques introduced in the stage are related to previously discussed techniques, and the process is repeated. In general, Section \ref{sec: review_select_steps} discusses ideas on how a hint system can select next-steps from a pool of existing steps. Section \ref{sec: review_generate_steps} then extends upon this by discussing how next-steps can be automatically generated using a goal. Following this, Section \ref{sec: review_features} investigates the uses of program features, including how direct and indirect comparisons can be used to produce hints, and how features can be attached to pre-written teacher feedback. Finally, Section \ref{sec: review_repair} discusses ideas for automatically repairing programs in order to produce hints. In the next section (Section \ref{sec: insights}), insights gained through this review will be presented, including a single, coherent picture of all of these ideas.

It is important to note that these sections do \textit{not} represent a general categorisation of hint techniques. Ideas from the different sections are related and can overlap. The intended purpose is to introduce new ideas and integrate these into a bigger picture as the survey progresses.

\subsection{Selecting Next Steps from Peer or Teacher Program States} \label{sec: review_select_steps}

A number of hint systems utilise existing data, such as peer actions or teacher solutions, to produce hints for help-seeking students. This survey begins by considering a subset of such systems, which convert existing data into program states, then select from these the most appropriate next state for the current student. While the selected next state may be given directly to students as hints, note that some systems, such as ITAP \cite{Rivers-K:2017}, involve additional steps before or after the next state is selected, and these further steps will be discussed in later sections.

\subsubsection{Example - Hint Factory.} 
\begin{figure}[h!]
	\centering
	\captionsetup{singlelinecheck=off}
	\includegraphics[width=0.9\linewidth]{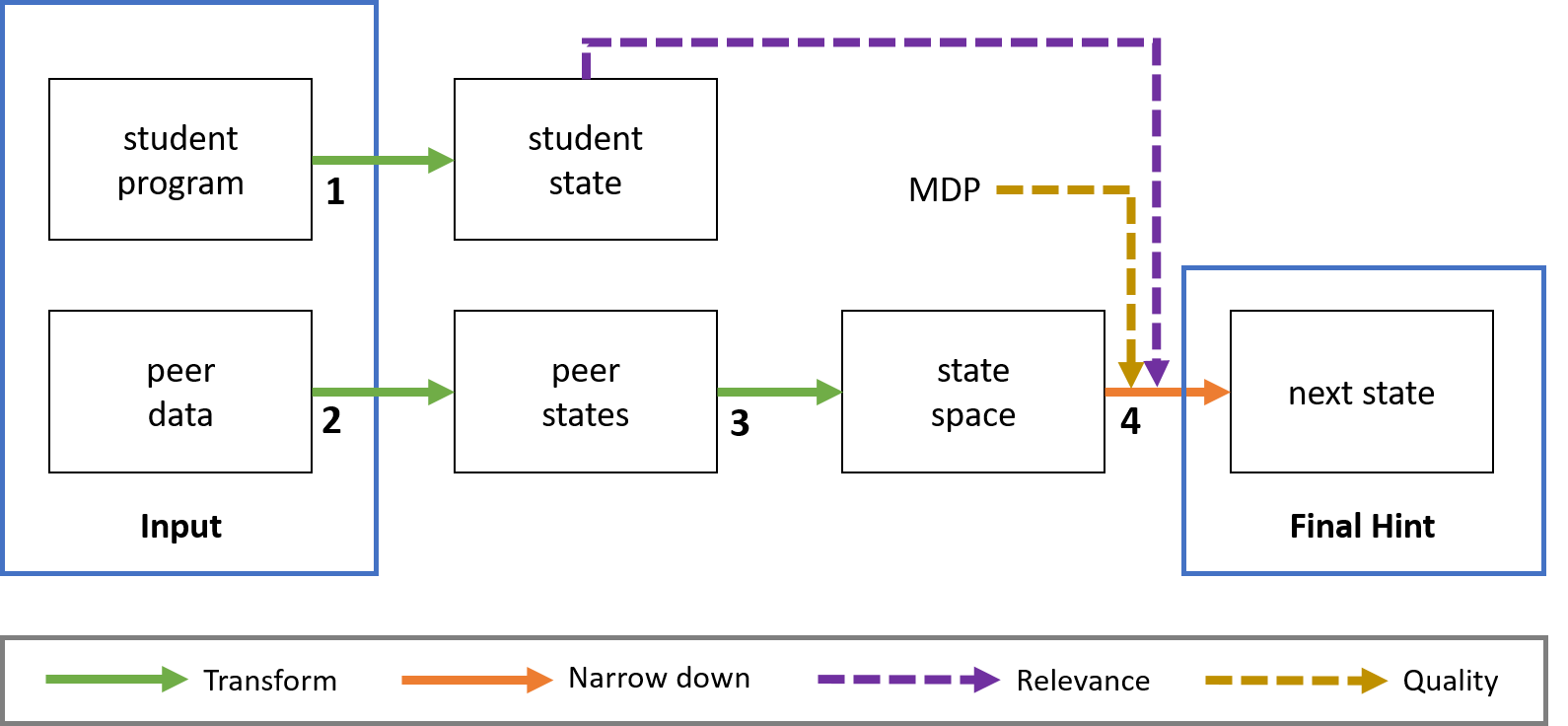}
	\caption[test]{A visualisation of the Hint Factory technique, using the HINTS framework (the colours correspond to the colours in Figure \ref{fig:framework}). \textbf{Input}: a student's program and peer data (i.e. programs submitted by peers). \textbf{Output hint}: the best next state for the student's program. \textbf{Processes}:	}
	\begin{enumerate}
		\item The student's program is \underline{transformed} into some general state.
		\item  Similarly to (1), peer programs are \underline{transformed} into states. (Note that, for efficiency purposes, this step is usually done before a student requests a hint, but could also be done after. In this paper, we focus on the automated process rather than the timing).
		\item The peer states are \underline{transformed} into a state space.
		\item The set of all states in the state space are \underline{narrowed down} to the best next one for the current student. This best state is chosen using a \underline{relevance} criterion - the next step must be reachable from the student's current state in the network - and also a \underline{quality} criterion - it must be the best next step from that position as determined by a  Markov Decision Process (MDP). 
	\end{enumerate}
	\label{fig:generating-hints-intermediate-output-diagram}
\end{figure}

Originally presented in the context of a logic tutor \cite{Stamper:BLC:2008} and later adapted to the domain of programming \cite{Iii-HB:2014,Hicks-PB:2014,Price-DB:2016}, the Hint Factory is a technique that uses peer data and a Markov Decision Process (MDP) to produce next-step hints for students. To produce hints, program submissions made by peers are transformed into a state space where each state represents a particular class of programs, and edges connecting the states represent how students transition between them (thereby indicating paths to various solutions). A student's submission is then matched to a state in this state space, and the MDP is used to determine the best next state for that student. Figure \ref{fig:generating-hints-intermediate-output-diagram} shows the general steps involved in the Hint Factory approach, and how these fit into the HINTS framework. 

\subsubsection{Discussion}
One key decision when producing a hint system such as this is how to select the best next state. In the Hint Factory approach, an MDP was used, but other options have also been explored in recent work. These include finding the most common next-state of peers in a similar position \cite{Chow-YKC:2017} (``Code-based" technique), using a scoring function that accounts for popularity, correctness and distance from the student's current state \cite{Rivers-K:2014}, finding the path with the shortest expected time to a solution \cite{Piech-SHG:2015} (``Poisson Path"), finding the most likely path an average student would take from the current state \cite{Piech-SHG:2015}(``Independent Probable Path") or using a custom distance measure to select the closest state \cite{Price-ZB:2017} (SourceCheck, solution matching step). With respect to the HINTS framework, each of these selection techniques could replace the MDP in Figure \ref{fig:generating-hints-intermediate-output-diagram} as the quality criterion. Note that the selected next state can be a combination of existing states, rather than a directly existing one, as in the Continuous Hint Factory \cite{Paassen-HPBGP:2017}, where the weighted sum of peer edits is used to select the best next state.

While there is currently no definitive evidence to suggest which of the these techniques are most effective at producing hints, some comparison studies have been conducted. In \cite{Piech-SHG:2015}, the authors' suggested ``Poisson Path" and ``Independent Probable Path" techniques most closely matched next-steps chosen by human experts. However, the programming exercises tested were simple and there was variation in technique performance across exercises. In addition, agreement or disagreement with experts does not necessarily imply high or low quality hints. These types of techniques were also compared in \cite{Price-DZPLCB:2019}.

Another key decision when producing a hint system such as this is how to represent student programs as states. In recent work, many variations have been presented. These include representing programs as ASTs \cite{Chow-YKC:2017}, based on their output \cite{Iii-HB:2014,Hicks-PB:2014} (``worldstates"), based on a standardised syntactic form, known as a \textit{canonical form} \cite{Iii-HB:2014} (``codestates") or based on their components  \cite{Price-DB:2016} \footnote{Note that in this case each component of a student's program is used to represent it in a different state space, so multiple next steps can be generated on each of the parts.} (``root paths"). Programs may also be represented by the inputs they were tested on by students \cite{Chow-YKC:2017} (input hints), or as points in a continuous space \cite{Paassen-HPBGP:2017}. Such representations often govern the nature of the hint given to students (e.g. if output is used, the the hints would suggest the next output to aim for. If input is used, the hints might suggest the next input to try testing the program on). Each of these representation steps are considered to be transformations under HINTS, and would replace steps 1 and 2 in Figure \ref{fig:generating-hints-intermediate-output-diagram}. 

Since there can be large variations in the programs written by students, using general states to represent these programs can help to reduce the size of the state space, and increase the probability of a match being found. However, the more general the states become, the less information there is available in the next-step. For example, in \cite{Iii-HB:2014}, the ``worldstates" were more general than the ``codestates" (there was less variation in the output than the syntax), so the state space was smaller. However, knowing the next ``worldstate" would only give a student information on how the output should change, and not the syntax. While work such as this has provided insight into the relationship between state type and hint availability, it is still unclear how the balance between next step availability and next step information content impacts on the quality of hints. In any case, when deciding upon a representation, these factors should be considered.

\subsection{Generating Next Steps Towards a Goal} \label{sec: review_generate_steps}  

In the previous section, the discussed techniques involved narrowing down a set of existing program states to just the most appropriate ones for the current student. Here, the discussion is extended to systems that \textit{produce} their own next states, as opposed to selecting them, by working towards some \textit{goal} program. Note that there are other approaches to producing next steps that do not involve a goal program, but these will be discussed in later sections.

\subsubsection{Example - Program Strategies in AskElle.} 
\begin{figure}[h!]
	\centering
	\captionsetup{singlelinecheck=off}
	\includegraphics[width=0.9\linewidth]{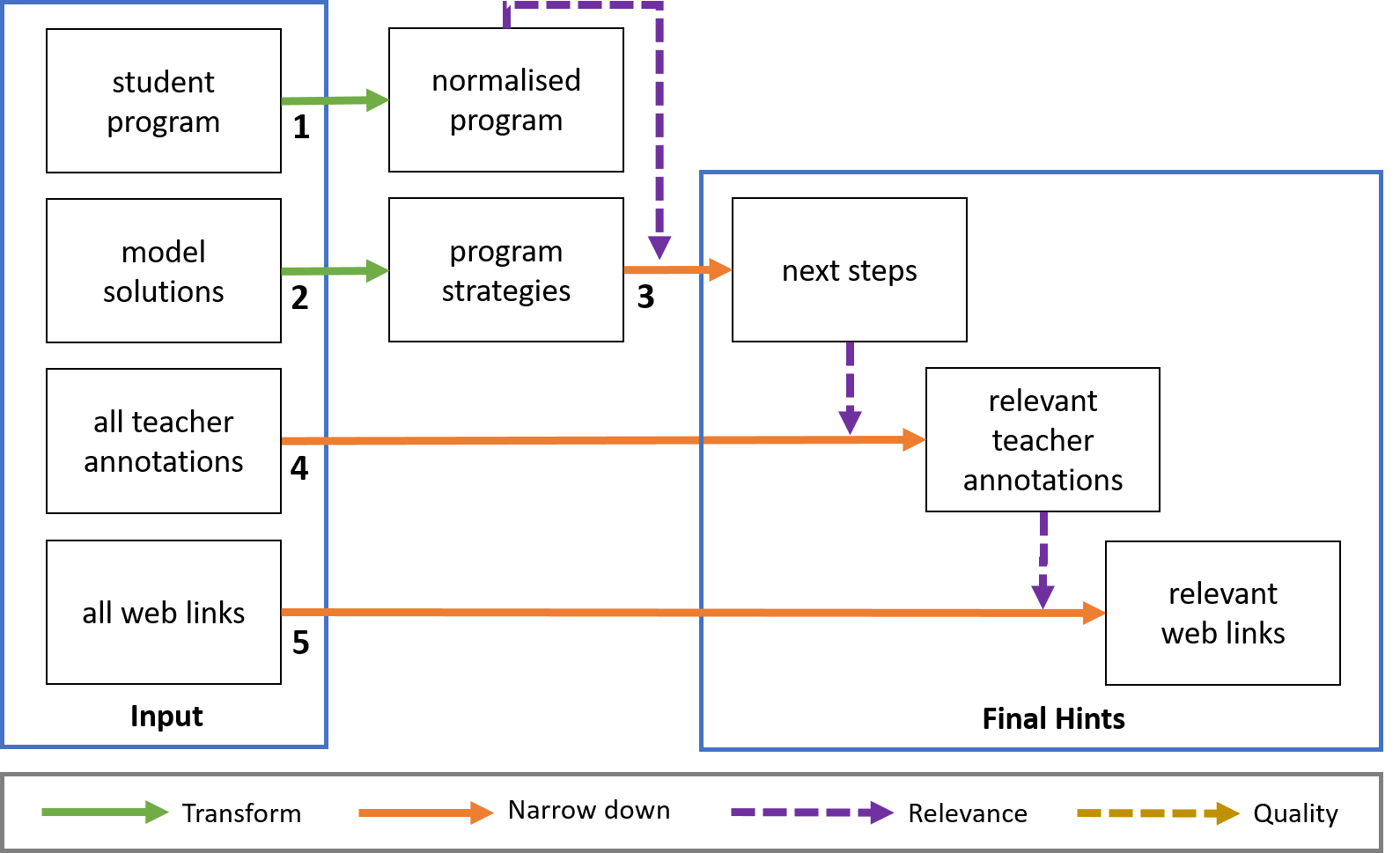}
	\caption[test]{A visualisation of how the program strategy hint technique in AskElle \cite{Gerdes-HJV:2017} fits into the HINTS framework. \textbf{Input}: a student's program, model solutions written by teachers, a set of annotations written by teachers to act as feedback and a set of possible links to documentation (whether explicit or implicit). \textbf{Output hints}: possible next steps a student can take, with relevant teacher annotations and links to documentation. \textbf{Processes}:
		\begin{enumerate}
			\item  A student's program is \underline{transformed} into a standardised form using transformations that affect syntax but not output.
			\item 	Model solutions are \underline{transformed} into a series of steps called \textit{program strategies} using a strategy language.
			\item  The program strategies are \underline{narrowed down} to just the next steps \underline{relevant} to the current student's normalised program. 
			\item Teacher annotations are \underline{narrowed down} to just the ones \underline{relevant} to the selected next steps
			\item the set of all possible web links are \underline{narrowed down} based on a \underline{relevance} criterion - that they link to documentation about prelude functions appearing in the selected teacher annotations.
		\end{enumerate}	
	}
	\label{fig:askelleflow}
\end{figure}

In \cite{Gerdes-HJV:2017}, the authors present a Haskell programming tutor, AskElle, which produces automated hints by using model solutions to a programming exercise. In particular, it converts sets of model solutions written by teachers into steps leading to these solutions, called \textit{programming strategies}. This is achieved through the use of a strategy language defined by the authors, which specifies how parts of a solution may be built up from others. The generated steps are in the form of a context free grammar, and can be used to parse a student's program to find potential next steps. Note that the student programs are \textit{normalised} (converted to a standard form) to increase the chance of matching the program to a step. Once potential next steps are found, teacher annotations associated with those steps are also given as hints, and any functions appearing in these annotations are automatically linked to external web pages with further information. Figure \ref{fig:askelleflow} shows how this hint method fits into the HINTS framework. 

Note that, in this example, three different types of hints are produced - next steps, relevant teacher annotations and relevant web links - as shown in Figure \ref{fig:askelleflow}. These have all been included for completeness, but the next steps component is focus of this section. The ideas behind the remaining components will be discussed later in Section \ref{sec: review_features}.

Also note that AskElle employs a second separate technique for generating hints, property-based testing, which will also be discussed in Section \ref{sec: review_features}.

\subsubsection{Discussion}
Notice firstly that AskElle produces its own steps from model solutions, instead of using existing steps. Similar approaches to this that use model solutions include \cite{Keuning-HJ:2014}, which also uses program strategies, and AutoTeach \cite{Antonucci-ENPM:2015}, where a solution is automatically converted into steps of increasing detail, called \textit{hint levels}. The process for generating hints in \cite{Keuning-HJ:2014} fits into HINTS in the same way as the strategy technique in AskElle, but the details of the individual steps are different, since the hints are generated for different programming languages, and the web link step is omitted. AutoTeach, on the other hand, is quite different, because the student program is not used in the hint generation process at all. Instead, the steps towards the solution are given incrementally to students as they request more hints. In addition to this, the details of how model solutions are converted into steps are different - solution steps are generated using customisable visibility rules defined by the authors instead of a strategy language, and there is only one path to one solution. In all of these systems, however, the result is a series of steps that can lead from a blank submission to correct programs.

While AskElle, AutoTeach and \cite{Keuning-HJ:2014} generate the set of hint steps in advance, another option is to wait until the student's program is known, then find the direct edits between that program and a goal program. This is done in the ``edit inference" step of SourceCheck \cite{Price-ZB:2017}, where the edits include deletion, movement, reordering and insertion of code to direct students to a solution. Note that this example was discussed previously to motivate the framework, so a description of how it fits into the framework can be seen in Table \ref{tab:step_examples} in Section \ref{sec: themes in hint generation}. A similar process is also involved in the hint generation process of ITAP \cite{Rivers-K:2017} and \cite{Sharma-AMK:2018} where, after the closest solution has been found, edits between this solution and the student's are generated and then, in the case of ITAP, further processed to produce next-steps.

One advantage of pre-computing steps to a solution is that these steps can be more easily attached to teacher-authored hints, since they are known in advance. Indeed, this is the case in all three of AskElle, AutoTeach and \cite{Keuning-HJ:2014}. However, a disadvantage is that it can be more difficult to deal with mistakes - if a student makes a mistake, they may no longer be on a path to a solution, meaning a next-step hint cannot be produced using that solution. It is possible to adapt the technique by including buggy solutions, so that the system can recognise mistakes if a student is on a path to these \cite{Gerdes:2012}, but this means the teacher must anticipate the types of mistakes students will make. Edit based approaches can still produce next-steps if there are mistakes by simply identifying parts of the solution not in the student's program.

One advantage of these techniques over the ones in the previous section is that they do not require existing steps. However, this also makes them sensitive to the input model solutions. In particular, if a student's attempt is completely different from any known  model solutions, then hints may either be unavailable or of low quality. For example, in the worst case, an edit-based approach might instruct a student to delete everything then re-write the solution. The program strategies approach might not be able to match a student program to a step towards a known solution, leaving no next-step hint. As such, these approaches are most effective when the model solutions are reflective of the different strategies students can take to solve a problem.

It is interesting to note that, when model solutions are not sufficiently close to the student's program, teachers using these systems must consider a trade-off between hint quality and hint availability. By setting restrictions on when hints can be produced (e.g. only when the model solution is close enough or only when the hint increases the number of tests passed, as in ITAP), teachers can increase the average relevance or usefulness of hints, but this will also reduce their availability. Conversely, if restrictions are removed, hint availability can be increased, but at the cost of quality. For example, if a student's program cannot be matched on a path to a solution using the program strategies approach, they can still be given a worked example to some solution - just perhaps not the one they were aiming for. Since the purpose of a hint system is to produce hints, high hint availability is clearly desirable. However, some recent research has suggested that poor quality hints can discourage students from seeking further help \cite{Price-ZB:2017-2}, so the balance between hint availability and quality in these systems is an important consideration that requires further investigation.

Note that the step generation techniques discussed in this section can be used in combination with the step selection techniques discussed in the previous section. Specifically, after a set of steps is automatically generated, an appropriate next step can be selected from these, as in AskElle. In addition, after an appropriate next step is selected, the difference between this state and the student's current state may be too large, so the automated generation techniques can use this as a goal to automatically generate smaller steps. A good example of how these ideas can be combined is ITAP. Here the nearest solution is first selected. This solution is then changed slightly to better match the student, using an idea discussed later in Section \ref{sec: review_repair}. Finally, smaller ``micro" edits are \textit{generated} towards the solution.

\subsection{Comparing Program Features} \label{sec: review_features}

So far, we have explored two important hint technique ideas - the selection of next steps from past data, and the generation of next steps using a goal. We now turn to another important idea in hint generation - the utilisation of \textit{program features}, such as output or structure, to produce hints. Specifically, we discuss how features of a student program can be compared to expected features to produce hints based on dissimilarity, or to assign pre-written teacher feedback to the student. Note that these are frequently paired with ideas from other sections to produce complex and interesting hint techniques.

\subsubsection{Example - Codewebs Engine}
\begin{figure}[h!]
	\centering
	\captionsetup{singlelinecheck=off}
	\includegraphics[width=0.9\linewidth]{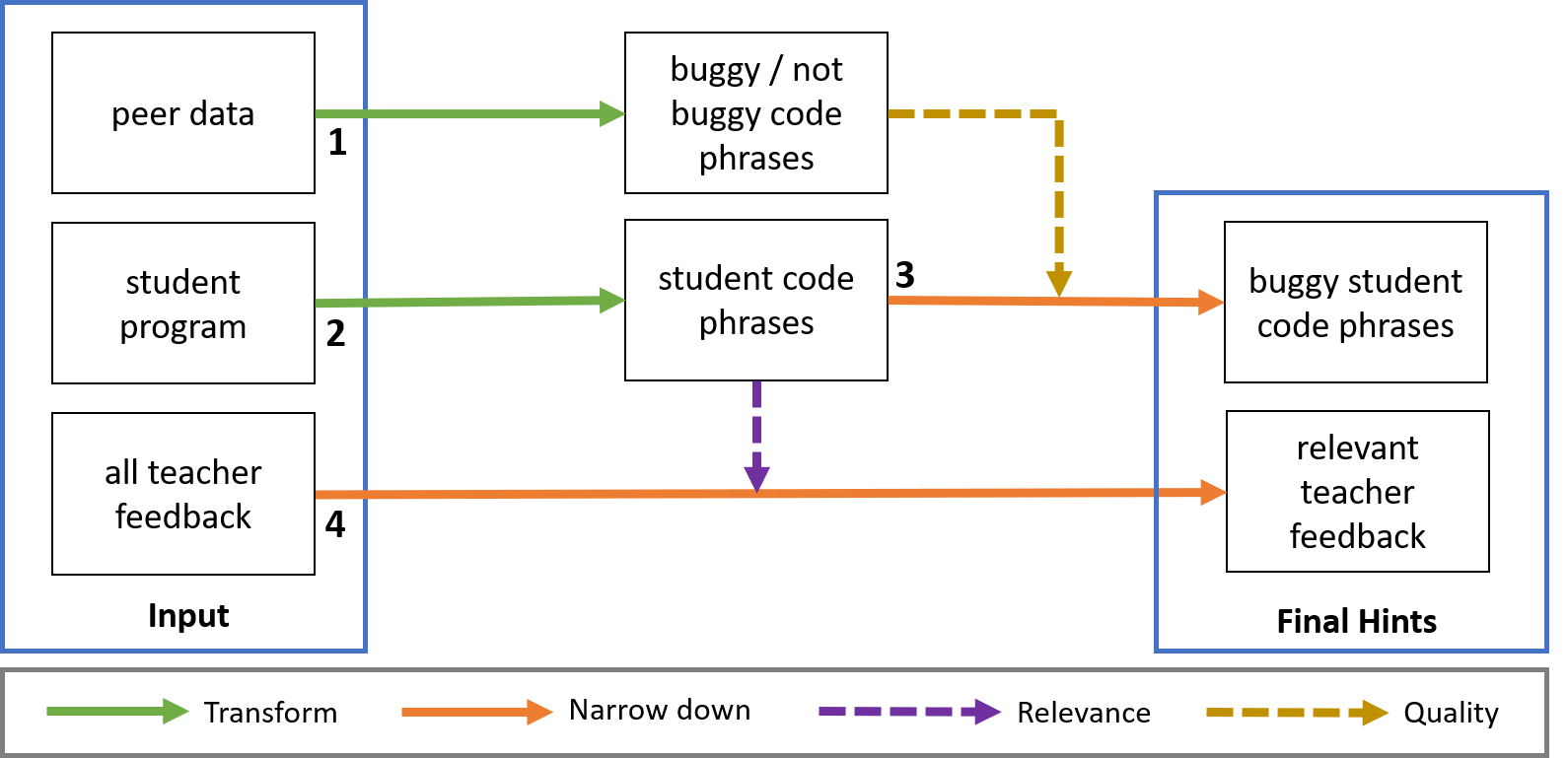}
	\caption[test]{A visualisation of how the hint techniques in Codewebs \cite{Nguyen-PHG:2014} fits into the HINTS framework. \textbf{Input}: peer data, a student's program and teacher feedback to be triggered by particular program pieces (code phrases). \textbf{Output hints}: locations of potential bugs in student programs and relevant teacher feedback for the student's program. \textbf{Processes}:
		\begin{enumerate}
			\item Peer programs are \underline{transformed} into a set of buggy and non-buggy code phrases. This is done by transforming all peer programs into ASTs and then, using a technique for discovering probabilistic equivalence classes presented by the authors, collecting similar subtrees together. Specific code phrases are then identified as buggy or not based on their association with incorrect or correct peer programs.
			\item  Similarly to the peer programs, a student's program is \underline{transformed} into a set of code phrases.
			\item 	The student code phrases are \textit{narrowed-down} to specific buggy phrases based on a \underline{quality} criterion -  that they have been identified as buggy in the peer data.
			\item Teacher feedback is \underline{narrowed down} to just the feedback \underline{relevant} to the student code phrases.
		\end{enumerate}	
	}
	\label{fig:codewebs}
\end{figure}

In \cite{Nguyen-PHG:2014}, the authors present a technique for extracting patterns, called \textit{code phrases}, from large numbers of peer programs. They then show how these patterns can be used to identify bugs and scale-up teacher feedback to provide hints to new students. Specifically, they show how particular patterns can be attached to teacher feedback, or automatically identified as buggy by considering their relative frequency in incorrect peer programs. When these patterns are then identified in a new student's program, the teacher feedback or a bug warning can then be automatically offered to the student. A visualisation of how this technique fits into HINTS is given in Figure \ref{fig:codewebs}.

\subsubsection{Discussion}
Notice in this example that program features, in the form of code phrases, were used to generate hints. Program features can include anything directly derivable from the program, such as its output, intermediate states or syntactic patterns. They can also be different representations of the program, such as its AST, dependence graph or canonical (standardised) form. In addition, they can be more abstract features, such as which next step is best for the program, or which transformations correct it.

When discussing how program features can be used to generate hints, it is useful to consider one of the simplest approaches first. Specifically, a list of expected features can be compared to a list of actual features of a student's program, and the differences can then be highlighted as hints. This is the idea behind the very widely used feedback approach for programming \cite{Keuning-JH:2016}, test cases, where the student's output is compared to the expected output to find differences. Further examples are given in Table \ref{table: examples_comparing_features}. 

\begin{table}[h!]
	\begin{minipage}{\columnwidth}
		\begin{center}
			\caption{Examples of how features of a student program can be directly compared to a set of expected features in order to produce hints}
			\label{table: examples_comparing_features}
			\begin{tabular}{p{1.5cm}|p{12cm}}
				\toprule
				\textbf{Feature} & \textbf{Comparison Example(s)}  \\ 
				\hline 
				\hline
				Concepts  & In \cite{Chow-YKC:2017}, the ``concepts" present in a student's program (e.g. if statements, for loops) are compared to  expected concepts (i.e. the concepts frequently present in correct peer submissions). Any expected concepts missing from the student's program are then given as \textit{concept hints}. \\ 
				\hline 
				Properties & In the property-based testing technique in \cite{Jeuring-BGH:2014,Gerdes-HJV:2017} (AskElle) the properties of a student's program are compared to the properties it is expected to satisfy. Information about any unsatisfied properties is then given as hints.  \\ 
				\hline 
				Syntax &   In \cite{Price-ZB:2017} (SourceCheck) and \cite{Rivers-K:2017} (ITAP) the syntax of a student's program is compared to the expected syntax (model solution). The differences are then used to produce hints in the form of syntactic edits. Interestingly, since these techniques can also be viewed as guiding a student towards a goal (see Sec. \ref{sec: review_generate_steps}), they are good examples of how hint techniques can be viewed from different perspectives, and how different hint ideas can relate.\\ 
				\bottomrule 
			\end{tabular} 
		\end{center}
		\bigskip\centering		
	\end{minipage}
\end{table}

Extending upon the idea of comparing features to a set of correct features, sometimes features are only expected under certain conditions. This is a key idea behind constraint-based tutoring systems \cite{M:2012}, which check student programs against a set of \textit{constraints}. These constraints involve a \textit{satisfaction condition} (i.e. the expected features) and a \textit{relevance condition} (i.e. the condition under which these features are expected). In \cite{Marin-PSR:2017}, patterns are matched against student programs using constraints and dependence graphs to produce hints. 

Note that the comparison between features can vary in directness. For instance, in the Codewebs example, every potentially buggy pattern in a student's program is not reported directly to the student. Instead, the AST structure is considered so buggy patterns closer to the leaves take precedence over their ancestors. Similarly, in \cite{Edmison-E:2015} (spectrum-based fault localisation) and \cite{Paassen-JH:2016}, the student's output and execution trace is compared to the expected output and execution trace (respectively), and the differences are used to infer which parts of the program contain bugs. As such, the incorrect output and intermediate values are not directly reported, but rather the program parts associated with them. 

Notice that the Codewebs example also introduces a second approach to generating hints from features.  Namely, teacher feedback can be attached to particular features (e.g. code phrases), then automatically scaled-up to any students whose programs have these features. This is in contrast to the previously discussed techniques, where the features of a student's program were compared to expected features, either directly or indirectly. Some further examples of using features to scale-up teacher hints are given in Table \ref{table: examples_group_teacher_hints} along with the Codewebs example.

\begin{table}[h!]
	\begin{minipage}{\columnwidth}
		\begin{center}
			\caption{Examples of how teacher feedback can be attached to particular features of a student program in order to produce relevant hints.}
			\label{table: examples_group_teacher_hints}
			\begin{tabular}{p{1.5cm}|p{12cm}}
				\toprule
				\textbf{Feature} & \textbf{Teacher Hint Example(s)}  \\ 
				\hline 
				\hline
				Output & ViDA \cite{Lee-YTWP:2018}, CSF$^2$ \cite{Haldeman-TBBSYN:2018} - teacher hints are written for various incorrect outputs, or sets of output, on different test cases. When the output of a student's program matches one of these known cases, the corresponding hint is given to the student.\\
				\hline
				Corrections & MistakeBrowser \cite{Head-GSSFDH:2017} - teacher feedback is attached to clusters of peer programs which are corrected by the same transformations (see Sec. \ref{sec: review_repair} for details of how these are found). When a new student's program can also be corrected by one of these transformations, the teacher feedback for that cluster is given.\\
				\hline
				Strategy & \cite{Gulwani-RZ:2014} - teacher feedback is written for different algorithmic strategies. These strategies are identified based on the intermediate values of some expressions in the program. When a student submits an incorrect program following a particular strategy, the teacher feedback associated with that strategy is given. \\ \hline
				Patterns & Codewebs \cite{Nguyen-PHG:2014} - teacher feedback is written to address particular bugs in peer data. Then, when these bugs are identified in a new student's program through patterns (code phrases), the same teacher feedback can be given to the new student. \cite{Chen-NSN:2017} also suggests scaling-up teacher feedback using patterns from Codewebs.\\
				\bottomrule 
			\end{tabular} 
		\end{center}
		\bigskip\centering		
	\end{minipage}
\end{table}

If these teacher hint examples were represented diagrammatically under HINTS, then all examples would include Step 4 of Codewebs (see Figure \ref{fig:codewebs}), where teacher feedback is narrowed down to relevant teacher feedback based on some feature of the student's program. They would also either include a transformation step similar to Step 2 of Codewebs to represent the student program in terms of its features, or a series of steps to produce those features. For example, MistakeBrowser finds features (i.e. corrections) by first repairing the student program, using steps described in the next section. Note that the process of writing teacher hints for particular features is itself not automated. However, these techniques can still be considered automated by our definition in Section \ref{sec: scope} if the teacher hints are available, pre-written, as input. As such, this manual aspect is treated as an input through the HINTS framework, rather than an automated step.

Just as the comparison-based ideas could be indirect, note that techniques for scaling-up teacher feedback can also be indirect. For example, in \cite{Piech-HNPSG:2015}, teacher hints are used as supervised labels to train a classifier to predict correct hints from program features. This classifier can then be used to propagate hints to new programs. As such, hints are not directly generated from the program features, but instead from a model trained on the features.

While this section discusses techniques for producing hints by comparing features or attaching them to teacher hints, note that there are many other uses of program features. In fact, almost every hint technique uses program features in some way. For example, in Section \ref{sec: review_select_steps}, where the techniques involved selecting next steps, program features were often used to produce program states. The techniques discussed in the next section will also make use of program features. As such, even though this section discusses two specific applications of program feature extraction, the set of possibilities is not limited to these applications.

\subsection{Automatically Repairing Programs} \label{sec: review_repair}

In the previous sections, many interesting ideas for producing hints were discussed. These included selecting next steps, generating next steps using a goal, and also utilising the program features to produce hints. This section builds on these by considering another important idea. In particular, it considers techniques for automatically repairing student programs in order to to produce feedback.

\subsubsection{Example - SYNFIX}
In \cite{Bhatia-S:2016}, the authors present a method for automatically correcting syntactic errors in student programs using  machine learning and peer data. Specifically, the peer data is used to train a recurrent neural network (RNN) to predict correct sequences of program pieces (tokens). When a new student requires help, the RNN can then be used to find new or alternative tokens to insert near the error location in order to correct the program. Corrections found in this way can then be used to produce hints for the student.

\begin{figure}[h!]
	\centering
	\captionsetup{singlelinecheck=off}
	\includegraphics[width=0.9\linewidth]{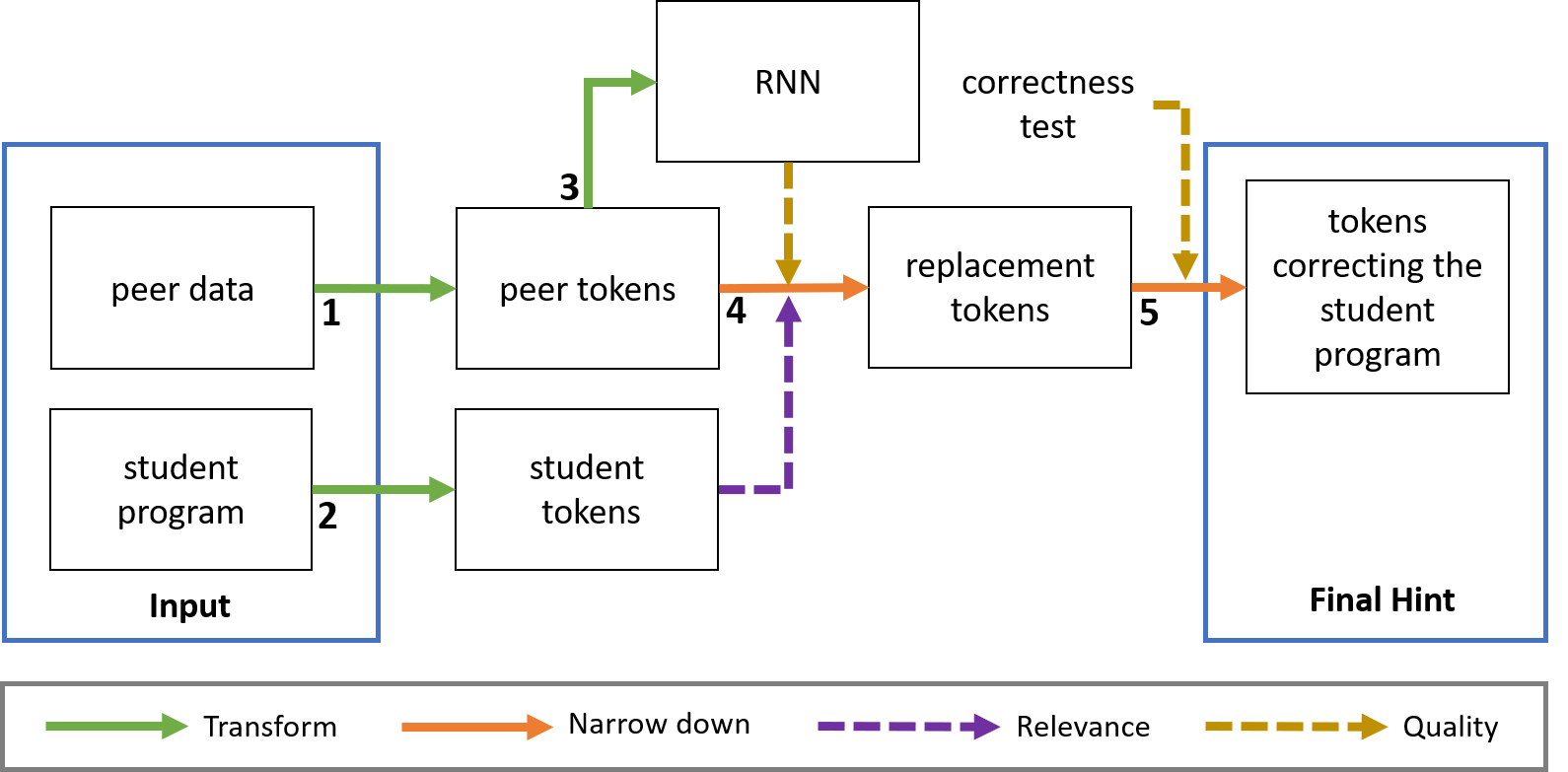}
	\caption[test]{A visualisation of how the program strategy hint technique in SYNFIX \cite{Bhatia-S:2016} fits into the HINTS framework. \textbf{Input}: peer data and a student's program. \textbf{Output hint}: token corrections to fix syntax errors in the student's program \textbf{Processes}:
		\begin{enumerate}
			\item Peer programs are \underline{transformed} into sequences of tokens, such as ``if", ``==", ``exp" etc.
			\item Similarly, a student's program is \underline{transformed} into a sequence of tokens
			\item  An RNN is trained on the sequences of correct peer tokens to learn correct patterns. As such, the correct peer tokens are \underline{transformed} into a model for  correct token sequences.
			\item  Peer tokens are \underline{narrowed down} to a series of replacement tokens based on a \underline{quality} criterion - that they are predicted by the RNN - and a \underline{relevance} criterion - that they follow on from other parts of the student program near the syntax error location.
			\item  The sequence of replacement tokens (which can be thought of as a set of subsequences) is \underline{narrowed down} to a single subsequence based on a \underline{quality} criterion - that it corrects the student program. 
		\end{enumerate}	
	}
	\label{fig:synfix}
\end{figure}

\subsubsection{Discussion}
In this example, first notice that there are two important steps. Firstly, the student programs are represented by their features (i.e. as sequences of tokens), similarly to the techniques in the previous section. Secondly, a model (RNN) is built using machine learning to represent how these features should be correctly combined (i.e. by learning correct token sequences). Similar techniques using machine learning are summarised in Table \ref{table: auto_repair_model} along with this one.

\begin{table}[h!]
	\begin{minipage}{\columnwidth}
		\begin{center}
			\caption{Examples of hint techniques that involve the use of machine learning to produce a model to represent correct program features}
			\label{table: auto_repair_model}
			\begin{tabular}{c|p{6.2cm}|p{4.3cm}}
				\toprule
				\textbf{Paper(s)} & \textbf{Model of Correct Features} & \textbf{Correction Technique}  \\ 
				\hline 
				\hline
				RLAssist \cite{Gupta-KS:2018} & Here, the program features are sequences of actions that correct the program. The model is an agent trained through reinforcement learning to learn correct sequences of actions. The learning process can be sped up by giving the agent examples of correct action sequences (``expert demos"). & The agent is used to produce actions that will correct the student's program.\\
				\hline
				\cite{Lazar-MB:2017} & The model is a set of rules for predicting correct and incorrect programs using AST patterns. In \cite{Lazar-MB:2017}, these rules are obtained by training a rule learner on correct and incorrect peer programs. In \cite{Movzina-LB:2018}, they are produced through argument-based machine learning (ABML). Note that, in the case of \cite{Movzina-LB:2018}, the model would be treated as input under HINTS since ABML is not automated)  & The rules are used to identify buggy AST patterns in the student's program, or good patterns to include in the program \\
				\hline
				SYNFIX \cite{Bhatia-S:2016} & An RNN, which is trained on sequences of correct peer tokens. & The RNN is used to find tokens that will correct the student's program near the error location. \\
				\bottomrule 
			\end{tabular} 
		\end{center}
		\bigskip\centering		
	\end{minipage}
\end{table}

It is interesting to note that these examples are highly related to the feature comparison techniques discussed in the previous section. Recall that the previous techniques involved comparing the features of a student's program to the expected features, either directly or indirectly. These examples are similar, but the comparison is even more indirect, because the expected features are now abstracted to a model first. For example, in SYNFIX, the sequences of tokens in a student's program are not directly compared to a list of expected token sequences, but they are input into a model which outputs expected token sequences.

As an alternative to the machine learning techniques discussed so far, which build a model to correct programs, another option is to perform a \textit{search} for possible corrections. This involves first defining some search space of possible program features (e.g. expressions or edits), and also some method for testing correctness (e.g. test cases). Then, starting with the student's current program, a search can be automatically performed over the search space to find edits that will satisfy the correctness test. Some examples of this are given in Table \ref{table: prog_synthesis}. Note that these search techniques relate to \textit{program synthesis} - the task of automatically creating a program that satisfies some criteria. The difference between the techniques from an educational context, compared to other areas, though, is that resources produced by teachers or peers can be utilised to correct programs.

\begin{table}[h!]
	\begin{minipage}{\columnwidth}
		\begin{center}
			\caption{Examples of hint techniques that involve a search to automatically correct student programs}
			\label{table: prog_synthesis}
			\begin{tabular}{c|p{5cm}|p{5.5cm}}
				\toprule
				\textbf{Paper} & \textbf{Search Space} & \textbf{Search Technique}  \\ 
				\hline 
				\hline
				\cite{Perelman-GG:2014}  & The search space is defined by a domain specific language (DSL), which defines the possible expressions. This is created by mining peer data for expressions that are used more than 10 times. &  The search begins at the student's program, which is slowly modified in steps. In each step, the current best program is modified to satisfy more input/output examples using the DSL. When all input/output examples are satisfied, the search is complete. \\
				\hline
				\cite{Lazar-SB:2017} & The search space consists of a set of rewrite rules learnt from past student data. Probabilities are assigned to each of these re-write rules based on their prevalence in the peer data & The search involves applying rewrite rules or combinations of them in order of increasing probability on the new student's program, until a correct solution is found or the system times out. In this way, more common solutions are explicitly favoured, and shorter solutions are implicitly favoured. \\
				\hline
				MistakeBrowser \cite{Head-GSSFDH:2017} & The search space consists of sets of transformations learned from peer program attempts. & The search involves trying different transformations until the program is corrected.\\
				\hline
				ITAP \cite{Rivers-K:2017} & (Note that this is a single part of a longer process). The search space comes from the closest known solution to a student's program. Specifically, the powerset (all possible subsets) of edits between this solution and the student's program is computed, and this forms the search space. & The search involves applying each of the subsets of edits to the student program, until the closest solution is found.\\
				\bottomrule 
			\end{tabular} 
		\end{center}
		\bigskip\centering		
	\end{minipage}
\end{table}

It is interesting to note a relationship between the two methods of correcting student programs discussed in this section, and the two methods of generating steps towards a goal discussed in Section \ref{sec: review_generate_steps}. Recall in Section \ref{sec: review_generate_steps} that steps could either be generated in advance before the student's program was seen, or they could be produced ``on the fly" by finding edits between the student's program and the goal. Here, again, there is a choice between producing a model of correct features in advance, or waiting until the student's program is known in order to perform a search. In some sense, the techniques are quite similar in nature, but in this section the goal is to pass all the test cases or successfully compile the program, and in Section \ref{sec: review_generate_steps}, the goal was to reach a model solution.

Once these techniques have been used to find corrections, there are many possible uses for these corrections when producing hints. For example, in MistakeBrowser \cite{Head-GSSFDH:2017}, the corrections are used to scale-up teacher hints. In \cite{Perelman-GG:2014} and \cite{Lazar-SB:2017}, the locations of the corrections are suggested to students as places to focus on. An interesting exploration of many different possible uses can be found in \cite{Suzuki-SGHDH:2017} where, for example, the authors suggest running the corrected code, then highlighting differences in the behaviour of this (i.e. the intermediate states) and the student's program.

Since all of these techniques aim to correct student programs, one important consideration when comparing these techniques is how many programs they are able to correct. It is difficult to directly compare these techniques in this regard, because they have been evaluated on different types of programs and serve different purposes (e.g. some correct syntax errors, and others logic errors). In addition, the types of evaluations performed on them have been different. For example, some authors evaluated how often the discovered corrections were later used by students. However, keeping in mind that the contexts were different, of the techniques that \textit{were} evaluated for the percentage of programs they could correct, the results varied greatly from around 27\% to 87\%. These figures are shown in Table \ref{table: correction_percent}. Note that, since less changes must be made to programs that are almost complete, one would expect these techniques to be most effective when a student's program is almost correct.

\begin{table}[h!]
	\begin{minipage}{\columnwidth}
		\begin{center}
			\caption{Reported percentage of programs able to be corrected through automatic repair techniques}
			\label{table: correction_percent}
			\begin{tabular}{c|p{3cm}|p{7.5cm}}
				\toprule
				\textbf{Paper} & \textbf{Correction \% (to the nearest \%)} & \textbf{Context - Types of Corrections Being Made} \\ 
				\hline 
				\hline
				MisktakeBrowser \cite{Head-GSSFDH:2017} & 87\% of students & Logic errors in simple programs. (e.g. "\textit{Repeated (720 students): takes as parameters a unary function f and a number n, and returns the nth application of f. For example, repeated(square,2)(5) returns square(square(5)), which evaluates to 625.}")\\
				\hline
				\cite{Perelman-GG:2014} & 65\% of attempts & Logic errors in programs written for a game, called ``Code-Hunt" \\
				\hline
				SYNFIX \cite{Bhatia-S:2016} & 32\% of programs, with partial corrections for an additional 6\%& Syntax errors in programs from an online introductory programming course\\
				\hline
				RLAssist \cite{Gupta-KS:2018} & 27\% of programs, 40\% of error messages & Syntax errors in programs from an introductory programming course \\
				\bottomrule 
			\end{tabular} 
		\end{center}
		\bigskip\centering		
	\end{minipage}
\end{table}

\section{Insights From Surveying Hint Techniques under the HINTS Framework} \label{sec: insights}
In the previous section, recent work on automated hints was surveyed in the context of HINTS. This section now presents some key insights resulting from this survey. In particular, it discusses how all of the previously introduced ideas can be integrated together into a single, coherent picture. In addition, it argues that the smaller components that comprise hint techniques should be considered when designing, communicating and evaluating hint systems. It explicitly demonstrates how viewing techniques in terms of their components can lead to important insights on connections between the field of automated programming hints and other areas, such as data-driven evaluation. Moreover, it explores how the similarities between hint techniques can offer insight into the nature of hint generation. Finally, it argues that the multitude of possible hints techniques producible from smaller steps is immense, and that this necessitates further work on evaluation methods.

\subsection{It is Possible to Fit Together Hint Techniques into a Coherent Picture Using their Components}
\begin{figure}[h!]
	\centering
	\label{big_picture}
	\captionsetup{singlelinecheck=off}
	\includegraphics[width=1\linewidth]{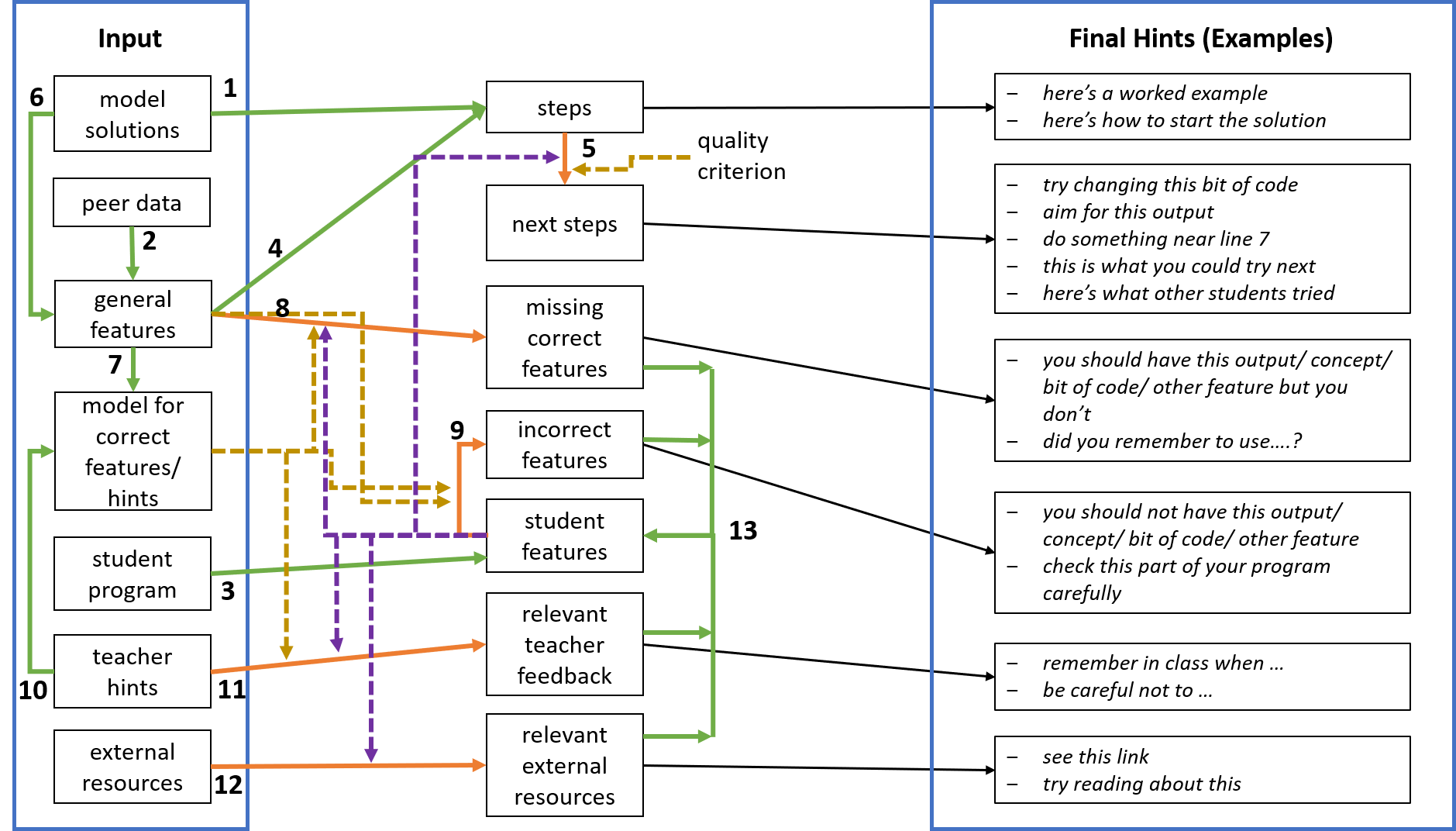}
	\caption[]{A general diagram summarising the main ideas of the surveyed papers and the relationships between them in the context of the HINTS framework. The leftmost blue box indicates types of hint data that can be given as input. The rightmost blue box gives examples of hints that can be produced from the indicated hint data. Green arrows denote transformation steps according to HINTS. Orange arrows indicate narrow-down steps, with any sources of quality or relevance criteria marked by gold or purple arrows respectively. Example techniques using each transformation and narrow-down step are given in the main text of this section, along with the sections in which they are discussed.}
	\label{fig:big-picture}
\end{figure}

After having reviewed some key ideas from hint generation techniques in Sections \ref{sec: review_select_steps} - \ref{sec: review_repair}, it is now possible to combine these ideas together into a single, coherent picture. This can be done by considering all of these techniques in the context of the HINTS framework - as a series of simpler steps - and then depicting similar steps together. By doing this, it is possible to gain insight into the current state of the field, including the relationships between existing techniques and the potential for future development. We show this general picture in Figure \ref{fig:big-picture}, with each of the steps described below.

\begin{enumerate}
	\item \textbf{model solutions $\rightarrow$ steps}. Model solutions can be transformed into steps in the form of program strategies (AskElle \cite{Gerdes-HJV:2017}, \cite{Keuning-HJ:2014}) or hint levels (AutoTeach \cite{Antonucci-ENPM:2015}). (See Sec. \ref{sec: review_generate_steps}).
	
	\item \textbf{peer data $\rightarrow$ general features}. Peer programs can be represented by their features for various purposes. For example, in order to form states in a state space, they can be represented by their AST form \cite{Chow-YKC:2017}, output \cite{Iii-HB:2014,Hicks-PB:2014}, canonical form \cite{Iii-HB:2014},components \cite{Price-DB:2016}, point in space \cite{Paassen-HPBGP:2017} or inputs they were tested on \cite{Chow-YKC:2017} (See Sec. \ref{sec: review_select_steps}). Also see Sec. \ref{sec: review_features} for further discussion of features.
	\item \textbf{student program $\rightarrow$ student features}. This is similar to (2), except the step is applied to the help-seeking student's program instead of peer programs.
	\item \textbf{general features $\rightarrow$ steps}. Program states can be collected together into a state space, as in \cite{Piech-SHG:2015,Chow-YKC:2017,Hicks-PB:2014,Price-DB:2016,Rivers-K:2014,Paassen-HPBGP:2017}. Note that if a technique involves steps (2) and (4), they can be treated as a single transformation step, depending on where the emphasis is being placed. (See Sec. \ref{sec: review_select_steps}).
	\item \textbf{steps $\rightarrow$ next steps}. A  set of potential next-steps can be narrowed down to just the ones relevant to the current student's program state. This can involve various quality criteria based on MDPs \cite{Iii-HB:2014,Hicks-PB:2014,Price-DB:2016} and other approaches \cite{Chow-YKC:2017,Rivers-K:2014,Piech-SHG:2015,Price-ZB:2017,Paassen-HPBGP:2017} (See Sec. \ref{sec: review_select_steps}). These can also come from program strategies \cite{Gerdes-HJV:2017,Keuning-HJ:2014} (See Sec. \ref{sec: review_generate_steps}).
	\item \textbf{model solutions $\rightarrow$ general features}. Model solutions can be divided into pieces to form a set of features (i.e. parts of programs).
	\item \textbf{general features $\rightarrow$ model for correct features}. A model can be built on the general features to learn which features are correct/ incorrect using, for example, reinforcement learning \cite{Gupta-KS:2018}, rules \cite{Lazar-MB:2017} or an RNN \cite{Bhatia-S:2016}. (See Sec. \ref{sec: review_repair}).
	\item \textbf{general features $\rightarrow$ missing correct features}. A set of general expected features can be compared to features of the student's program. Expected features not in the student's program can be identified as missing correct features. For example, these include edits to a solution in \cite{Price-ZB:2017,Rivers-K:2017,Sharma-AMK:2018} that involve adding code. (See Sec. \ref{sec: review_generate_steps}). It could also include missing concepts \cite{Chow-YKC:2017} or properties \cite{Jeuring-BGH:2014,Gerdes-HJV:2017} (See Sec. \ref{sec: review_features}). The set of features can also be searched until some correctness test is passed as in \cite{Perelman-GG:2014,Lazar-SB:2017,Head-GSSFDH:2017,Rivers-K:2017}, or a model of correct features can be used to narrow them down as in \cite{Gupta-KS:2018,Lazar-MB:2017,Bhatia-S:2016,Movzina-LB:2018}
	\item \textbf{student features $\rightarrow$ incorrect features}. Similarly to (8), student features can be compared to expected features, and the features in the student's program that are not expected can be identified as incorrect, either directly or indirectly. For example, edits for deleting code in \cite{Price-ZB:2017,Rivers-K:2017,Sharma-AMK:2018}. (See Sec. \ref{sec: review_generate_steps}). These can also be based on buggy patterns \cite{Nguyen-PHG:2014}, constraints \cite{Marin-PSR:2017}, output \cite{Edmison-E:2015} or execution traces \cite{Paassen-JH:2016}. (See Sec. \ref{sec: review_features})
	\item \textbf{teacher hints $\rightarrow$ model for correct hints}. Teacher hints attached to particular program features can be transformed into a model to predict correct hints, as in \cite{Piech-HNPSG:2015}. (See Sec. \ref{sec: review_features})
	\item \textbf{teacher hints $\rightarrow$ relevant teacher hints}. Teacher feedback can be narrowed down to just the feedback relevant to the student's features. These can include output \cite{Lee-YTWP:2018,Haldeman-TBBSYN:2018}, edits that correct the program \cite{Head-GSSFDH:2017} or strategy \cite{Gulwani-RZ:2014} (See Sec. \ref{sec: review_features}).
	\item \textbf{external resources $\rightarrow$ relevant external resources}. External resources, such as web links \cite{Gerdes-HJV:2017}, can be narrowed-down to just those relevant to the student's features (See Sec. \ref{sec: review_features}).
	\item \textbf{student features cycle}. Any next steps, missing correct features, etc. can be treated as features of the student program, which can be used to build more hints. 
\end{enumerate}

\subsection{During the Design, Communication and Evaluation of Hint Systems, the Smaller Components that Comprise Hint Techniques Should be Considered}
The fact that hint techniques are comprised of many smaller steps suggests that these steps should play a key role in the design, communication and evaluation of hint systems. In recent work, there have been many interesting techniques and ideas presented about automated hint generation. However, without considering these techniques as a series of smaller processes that can be modified and re-purposed, we can miss opportunities to integrate them and utilise them in new work. As such, when communicating and designing new techniques, we should consider the different steps comprising these techniques, the possible variations on these steps and whether these steps could be used in different contexts for different purposes. In addition, when evaluating hint systems, we should not only consider the effectiveness of entire techniques, but also evaluate the individual components and how choices of these affect the overall quality of hint systems.

\subsection{The Connection Between Hint Generation Techniques and Data-Driven Hint Evaluation Techniques Should Be Investigated Further}
Recently, along with the development of hint generation techniques, there has been much interest in hint \textit{evaluation} techniques. In particular, there has been a focus on \textit{data-driven} evaluation methods, which utilise historical student data to evaluate hint systems. Since evaluation methods serve a different purpose from hint generation techniques, one would expect these two kinds of techniques to be quite different. However, by considering hint techniques in the context of our framework, as a series of simpler components, it is possible to observe some interesting relationships between hint techniques and data-driven evaluation methods.

In order to see these relationships, first consider the narrow-down step of the HINTS framework. Recall that this step involves narrowing down a set of hint data to just the most appropriate subset for the current student, using relevance or quality criteria. As such, any hint technique with a narrow-down step is, in some sense, performing an evaluation of the hint data to decide what to select. For example, the SourceCheck \cite{Price-ZB:2017} hint technique (discussed previously) involves a narrow down step where all peer solutions are narrowed down to just the closest one. In order to do this, each of the potential solutions must be ``evaluated" for their quality, by checking their distance from the current student's program. By considering hint techniques in the context of HINTS, we can thus begin to notice a link between hint techniques and the general idea of evaluation.

Now consider an actual data-driven evaluation method, presented in \cite{Rivers-K:2017} to evaluate ITAP (which was introduced previously). This technique involves computing the number of edits needed to correct a student's program (i.e. its distance from the student's program). In essence, just as in SourceCheck, the solution is being evaluated based on how close it is to the student's program. While these techniques are not exactly the same, since the distance measure is different, this indicates there is 
an important relationship between this evaluation technique and the narrow-down step of SourceCheck.

We can observe another example of the correspondence between hint generation techniques and hint evaluation techniques by considering MistakeBrowser from \cite{Head-GSSFDH:2017}, and a different evaluation method for ITAP in \cite{Rivers-K:2017}. In MistakeBrowser, the hint system narrows down a large set of transformations to just a small number by ``evaluating" whether or not they are able to correct a student's program. Similarly, in \cite{Rivers-K:2017}, in addition to path length, there is an evaluation to check whether a series of hints (in the form of edits, which can be thought of as transformations) would actually lead students to a solution. As such, in both cases, transformations are evaluated based on whether they correct the student's program, suggesting an important link between these techniques.

Note that it is not only narrow-down steps that seem to correspond to evaluation techniques, but also transformation steps. Recall that these steps involve changing the representation of the data in order to produce hints. For example, in \cite{Iii-HB:2014}, which uses the HintFactory approach (see Sec \ref{sec: review_select_steps}), programs are transformed into states (i.e. \textit{worldstates}) based on their output, which are then transformed into a state-space. This relates to the evaluation method presented in \cite{Mcbroom-YKC:2018}, where peer programs are clustered and transformed into a state-space so that teachers can visualise how students complete an exercise. Two additional examples are described in Table \ref{hint_eval_vs_gen}, along with a summary of the examples already discussed. 

\begin{table}[h!]
	\begin{minipage}{\columnwidth}
		\begin{center}
			\caption{Links between steps in hint generation and data-driven evaluation techniques}
			\label{hint_eval_vs_gen}
			\begin{tabular}{p{7cm}p{7cm}}
				\toprule
				\textbf{Evaluation  Technique} & \textbf{Related Transformation or Narrow Down Step in a Hint Technique} \\
				\hline \hline
				check the number of edits between the student's program and the solution (i.e. the distance) (ITAP \cite{Rivers-K:2017}) & narrow down the set of all solutions by checking their distance from the student's program and choosing the closest one (SourceCheck \cite{Price-ZB:2017})\\
				\hline
				check how many chains of hints (transformations) actually lead to a solution (ITAP \cite{Rivers-K:2017}) & narrow down sets of transformations to just the ones that lead to a solution \cite{Head-GSSFDH:2017} \\
				\hline
				transform student programs into a standardised form and create an state space so teachers can visualise how students complete an exercise \cite{Mcbroom-YKC:2018} & transform student programs into a standardised form and create an state space so the next state can be selected \cite{Iii-HB:2014}\\
				\hline
				transform student programs into code-phrases then organise them so teachers can understand the data (e.g. they can ``count the number of students who submitted the same or a similar class of solutions") (Codewebs \cite{Nguyen-PHG:2014}) & transform student programs into code-phrases then organise them so automated hints can be given to students based on the code-phrases (Codewebs \cite{Nguyen-PHG:2014})\\
				\hline
				use a templating language to express model solutions in many different forms to account for different programming strategies when evaluating student solutions against experts.\cite{Price-ZDLB:2018}& use a templating language (``strategy language") to express paths to model solutions in many different forms to account for different programming strategies when guiding students to a solutions (AskElle \cite{Gerdes-HJV:2017})\\
				\hline
				\bottomrule
			\end{tabular}
		\end{center}
		\bigskip\centering		
	\end{minipage}
\end{table}

The link between transformation steps and data-driven evaluation techniques is perhaps not so surprising, considering that these evaluation techniques must convey information about the hint system to a teacher in an understandable way. This will often involve techniques for automatically representing large volumes of data in a coherent way, or transforming them into more accessible forms. Since this is also often the purpose of transformation steps in hint techniques, this suggests a correspondence between them.

While a full survey of evaluation methods is beyond this scope of this paper, there is clearly an important link between hint generation and data-driven evaluation methods, which would be a worthwhile avenue for further investigation. Perhaps there is not only potential to connect different hint techniques, but also to use their steps to improve evaluation methods, or to use the steps of evaluation methods to extend hint techniques.

\subsection{The Design of Hint Techniques Can Provide Insight into the Nature of Hint Generation}
It is clear from the HINTS framework that all automated hint techniques exhibit remarkable similarities in structure and in the processes used to construct them. In particular, they may all be described by two simple operations applied iteratively: narrowing down and transforming hints from previous levels. Considering their diversity and sophistication, this is surprising, and prompts the question of why this is the case.

One potential reason could be that the initial inputs to these systems are already highly complex, so a simple system could still leverage these to produce intelligent feedback. Indeed, hints or tests pre-written by teachers and work produced by other students contain vast amounts of information, and a system that narrowed this down could produce high quality hints. This could explain why even the simplest automated hint techniques can produce highly sophisticated hints.

Another potential reason for why hint techniques share such similar operations could be that these operations reflect the nature of programming hints. Programming hints often involve highlighting mistakes to students, so we would expect the system to have some way of narrowing down the student's program to just the mistakes. In addition, programming hints often involve suggesting new ways to proceed, so we would expect the system to have some knowledge about a goal, then to narrow this down to the parts the student has not yet succeeded in. In addition, programming hints can related to many different aspects of a student's program, such as output, style, structure, syntactic patterns or run-time, so we would also expect the system to need transformations to give feedback on these different aspects. As such, perhaps a reason why hint techniques involve transformation and narrow-down steps could be that these steps reflect the nature of programming hints. Perhaps this insight can help to shape the way we think and communicate about hint techniques, and motivate further work. For example, it would be interesting in future to investigate whether the HINTS framework could be extended to cover techniques from other domains, such as physics, logic or mathematics, where some hints could be of a similar nature.

\subsection{The Multitude of Possible Hint Techniques from Components Necessitates Further Work on Evaluation Methods}
From the survey of hint techniques presented in Section \ref{sec: example guided review}, it is clear that there are many different approaches to generating hints. Specifically, there are many different interchangeable and stackable components that comprise hint techniques, and different combinations of these can result in a vast number of possible hint approaches. This suggests a need to develop scalable and versatile evaluation methods that can cope with such a multitude of potential techniques.

Even now, with a relatively small number of techniques compared to the possible number in future, these techniques are so numerous that it is difficult to decide which are most effective for different scenarios (e.g. different programming languages, learners or courses). This is exacerbated by the fact that evaluation methods are often applied inconsistently \cite{Keuning-JH:2018}. Considering the potential for far more hint techniques in future, it is thus of great importance that we continue to develop and extend evaluation methods to ensure the full potential of automated hint techniques.

\section{Conclusion}
In this paper, we have surveyed and developed key theoretical ideas behind recent work from 2014-2018 on generating automated hints for programming exercises. Specifically, we have presented a novel framework, the HINTS framework, for describing techniques for hint generation, and surveyed these techniques in the context of this framework. We have shown that, by considering hint techniques as a series of smaller steps, it is possible to draw recent work together into a single coherent picture. We have argued that this perspective on hint techniques has implications for how we design, communicate and evaluate hint systems, and can provide useful insights into the nature of hint generation. Finally, we have identified a potential relationship between hint generation and evaluation techniques that could be utilised to improve both, and have argued that the piece-wise nature of hint techniques necessitates the further development of evaluation methods. By bringing more clarity to the area of automated programming hint generation, this work acts as an important step towards realising the full potential of automated programming tutors, with the ultimate goal of maximising educational outcomes.


\begin{thebibliography}{10}
	
	\bibitem{Head-GSSFDH:2017}
	A.~Head, E.~Glassman, G.~Soares, R.~Suzuki, L.~Figueredo, L.~D'Antoni, and
	B.~Hartmann, ``Writing reusable code feedback at scale with mixed-initiative
	program synthesis,'' in {\em Proceedings of the Fourth (2017) ACM Conference
		on Learning@ Scale}, pp.~89--98, ACM, 2017.
	
	\bibitem{Lee-YTWP:2018}
	V.~C. Lee, Y.-T. Yu, C.~M. Tang, T.-L. Wong, and C.~K. Poon, ``Vida: A virtual
	debugging advisor for supporting learning in computer programming courses,''
	{\em Journal of Computer Assisted Learning}, vol.~34, no.~3, pp.~243--258,
	2018.
	
	\bibitem{Haldeman-TBBSYN:2018}
	G.~Haldeman, A.~Tjang, M.~Babe{\c{s}}-Vroman, S.~Bartos, J.~Shah, D.~Yucht, and
	T.~D. Nguyen, ``Providing meaningful feedback for autograding of programming
	assignments,'' in {\em Proceedings of the 49th ACM Technical Symposium on
		Computer Science Education}, pp.~278--283, ACM, 2018.
	
	\bibitem{Lazar-MB:2017}
	T.~Lazar, M.~Mo{\v{z}}ina, and I.~Bratko, ``Automatic extraction of ast
	patterns for debugging student programs,'' in {\em International Conference
		on Artificial Intelligence in Education}, pp.~162--174, Springer, 2017.
	
	\bibitem{Iii-HB:2014}
	B.~Peddycord~Iii, A.~Hicks, and T.~Barnes, ``Generating hints for programming
	problems using intermediate output,'' in {\em Proceedings of the 7th
		International Conference on Educational Data Mining}, pp.~92--98,
	International Educational Data Mining Society, 2014.
	
	\bibitem{Lazar-SB:2017}
	T.~Lazar, A.~Sadikov, and I.~Bratko, ``Rewrite rules for debugging student
	programs in programming tutors,'' {\em IEEE Transactions on Learning
		Technologies}, vol.~11, no.~4, pp.~429--440, 2017.
	
	\bibitem{Edmison-E:2015}
	B.~Edmison and S.~H. Edwards, ``Applying spectrum-based fault localization to
	generate debugging suggestions for student programmers,'' in {\em Software
		Reliability Engineering Workshops (ISSREW), 2015 IEEE International Symposium
		on}, pp.~93--99, IEEE, 2015.
	
	\bibitem{Keuning-HJ:2014}
	H.~Keuning, B.~Heeren, and J.~Jeuring, ``Strategy-based feedback in a
	programming tutor,'' in {\em Proceedings of the Computer Science Education
		Research Conference}, pp.~43--54, ACM, 2014.
	
	\bibitem{Gerdes-HJV:2017}
	A.~Gerdes, B.~Heeren, J.~Jeuring, and L.~T. van Binsbergen, ``Ask-elle: an
	adaptable programming tutor for haskell giving automated feedback,'' {\em
		International Journal of Artificial Intelligence in Education}, vol.~27,
	no.~1, pp.~65--100, 2017.
	
	\bibitem{Price-ZB:2017}
	T.~W. Price, R.~Zhi, and T.~Barnes, ``Evaluation of a data-driven feedback
	algorithm for open-ended programming.,'' in {\em Proceedings of the 10th
		International Conference on Educational Data Mining}, pp.~192--197,
	International Educational Data Mining Society, 2017.
	
	\bibitem{Rivers-K:2017}
	K.~Rivers and K.~R. Koedinger, ``Data-driven hint generation in vast solution
	spaces: a self-improving python programming tutor,'' {\em International
		Journal of Artificial Intelligence in Education}, vol.~27, no.~1, pp.~37--64,
	2017.
	
	\bibitem{Piech-SHG:2015}
	C.~Piech, M.~Sahami, J.~Huang, and L.~Guibas, ``Autonomously generating hints
	by inferring problem solving policies,'' in {\em Proceedings of the Second
		(2015) ACM Conference on Learning@ Scale}, pp.~195--204, ACM, 2015.
	
	\bibitem{Price-DB:2016}
	T.~W. Price, Y.~Dong, and T.~Barnes, ``Generating data-driven hints for
	open-ended programming.,'' in {\em Proceedings of the 9th International
		Conference on Educational Data Mining}, pp.~191--198, International
	Educational Data Mining Society, 2016.
	
	\bibitem{Antonucci-ENPM:2015}
	P.~Antonucci, C.~Estler, D.~Nikoli{\'c}, M.~Piccioni, and B.~Meyer, ``An
	incremental hint system for automated programming assignments,'' in {\em
		Proceedings of the 2015 ACM Conference on Innovation and Technology in
		Computer Science Education}, pp.~320--325, ACM, 2015.
	
	\bibitem{Marin-PSR:2017}
	V.~J. Marin, T.~Pereira, S.~Sridharan, and C.~R. Rivero, ``Automated
	personalized feedback in introductory java programming moocs,'' in {\em Data
		Engineering (ICDE), 2017 IEEE 33rd International Conference on},
	pp.~1259--1270, IEEE, 2017.
	
	\bibitem{Chow-YKC:2017}
	S.~Chow, K.~Yacef, I.~Koprinska, and J.~Curran, ``Automated data-driven hints
	for computer programming students,'' in {\em Adjunct Publication of the 25th
		Conference on User Modeling, Adaptation and Personalization}, pp.~5--10, ACM,
	2017.
	
	\bibitem{Price-DZPLCB:2019}
	T.~W. Price, Y.~Dong, R.~Zhi, B.~Paa{\ss}en, N.~Lytle, V.~Catet{\'e}, and
	T.~Barnes, ``A comparison of the quality of data-driven programming hint
	generation algorithms,'' {\em International Journal of Artificial
		Intelligence in Education}, pp.~1--28, 2019.
	
	\bibitem{Le:2016}
	N.-T. Le, ``A classification of adaptive feedback in educational systems for
	programming,'' {\em Systems}, vol.~4, no.~2, p.~22, 2016.
	
	\bibitem{Striewe-G:2014}
	M.~Striewe and M.~Goedicke, ``A review of static analysis approaches for
	programming exercises,'' in {\em International Computer Assisted Assessment
		Conference}, pp.~100--113, Springer, 2014.
	
	\bibitem{Keuning-JH:2018}
	H.~Keuning, J.~Jeuring, and B.~Heeren, ``A systematic literature review of
	automated feedback generation for programming exercises,'' {\em ACM
		Transactions on Computing Education (TOCE)}, vol.~19, no.~1, p.~3, 2018.
	
	\bibitem{Keuning-JH:2016}
	H.~Keuning, J.~Jeuring, and B.~Heeren, ``Towards a systematic review of
	automated feedback generation for programming exercises,'' in {\em
		Proceedings of the 2016 ACM Conference on Innovation and Technology in
		Computer Science Education}, pp.~41--46, ACM, 2016.
	
	\bibitem{Ala-K:2005}
	K.~M. Ala-Mutka, ``A survey of automated assessment approaches for programming
	assignments,'' {\em Computer science education}, vol.~15, no.~2, pp.~83--102,
	2005.
	
	\bibitem{Ihantola-AKS:2010}
	P.~Ihantola, T.~Ahoniemi, V.~Karavirta, and O.~Sepp{\"a}l{\"a}, ``Review of
	recent systems for automatic assessment of programming assignments,'' in {\em
		Proceedings of the 10th Koli calling international conference on computing
		education research}, pp.~86--93, ACM, 2010.
	
	\bibitem{Le-SGP:2013}
	N.-T. Le, S.~Strickroth, S.~Gross, and N.~Pinkwart, ``A review of ai-supported
	tutoring approaches for learning programming,'' in {\em Advanced
		Computational Methods for Knowledge Engineering}, pp.~267--279, Springer,
	2013.
	
	\bibitem{Crow-LW:2018}
	T.~Crow, A.~Luxton-Reilly, and B.~Wuensche, ``Intelligent tutoring systems for
	programming education: a systematic review,'' in {\em Proceedings of the 20th
		Australasian Computing Education Conference}, pp.~53--62, ACM, 2018.
	
	\bibitem{Monperrus:2018}
	M.~Monperrus, ``Automatic software repair: a bibliography,'' {\em ACM Computing
		Surveys (CSUR)}, vol.~51, no.~1, p.~17, 2018.
	
	\bibitem{Silva:2011}
	J.~Silva, ``A survey on algorithmic debugging strategies,'' {\em Advances in
		engineering software}, vol.~42, no.~11, pp.~976--991, 2011.
	
	\bibitem{Tiam-S:2018}
	T.~J. Tiam-Lee and K.~Sumi, ``Adaptive feedback based on student emotion in a
	system for programming practice,'' in {\em International Conference on
		Intelligent Tutoring Systems}, pp.~243--255, Springer, 2018.
	
	\bibitem{Barron-ZHB:2015}
	M.~L. Barr{\'o}n-Estrada, R.~Zatarain-Cabada, F.~G. Hern{\'a}ndez, R.~O.
	Bustillos, and C.~A. Reyes-Garc{\'\i}a, ``An affective and cognitive tutoring
	system for learning programming,'' in {\em Mexican International Conference
		on Artificial Intelligence}, pp.~171--182, Springer, 2015.
	
	\bibitem{Annamaa-SV:2017}
	A.~Annamaa, R.~Suviste, and V.~Vene, ``Comparing different styles of automated
	feedback for programming exercises,'' in {\em Proceedings of the 17th Koli
		Calling Conference on Computing Education Research}, pp.~183--184, ACM, 2017.
	
	\bibitem{Stamper:BLC:2008}
	J.~Stamper, T.~Barnes, L.~Lehmann, and M.~Croy, ``The hint factory: Automatic
	generation of contextualized help for existing computer aided instruction,''
	in {\em Proceedings of the 9th International Conference on Intelligent
		Tutoring Systems Young Researchers Track}, pp.~71--78, 2008.
	
	\bibitem{Hicks-PB:2014}
	A.~Hicks, B.~Peddycord, and T.~Barnes, ``Building games to learn from their
	players: Generating hints in a serious game,'' in {\em International
		Conference on Intelligent Tutoring Systems}, pp.~312--317, Springer, 2014.
	
	\bibitem{Rivers-K:2014}
	K.~Rivers and K.~R. Koedinger, ``Automating hint generation with solution space
	path construction,'' in {\em International Conference on Intelligent Tutoring
		Systems}, pp.~329--339, Springer, 2014.
	
	\bibitem{Paassen-HPBGP:2017}
	B.~Paa{\ss}en, B.~Hammer, T.~W. Price, T.~Barnes, S.~Gross, and N.~Pinkwart,
	``The continuous hint factory-providing hints in vast and sparsely populated
	edit distance spaces,'' {\em arXiv preprint arXiv:1708.06564}, 2017.
	
	\bibitem{Sharma-AMK:2018}
	S.~Sharma, P.~Agarwal, P.~Mor, and A.~Karkare, ``Tipsc: tips and corrections
	for programming moocs,'' in {\em International Conference on Artificial
		Intelligence in Education}, pp.~322--326, Springer, 2018.
	
	\bibitem{Gerdes:2012}
	A.~Gerdes, {\em Ask-Elle: a Haskell Tutor}.
	\newblock PhD thesis, Universiteit Utrecht, 2012.
	
	\bibitem{Price-ZB:2017-2}
	T.~W. Price, R.~Zhi, and T.~Barnes, ``Hint generation under uncertainty: the
	effect of hint quality on help-seeking behavior,'' in {\em International
		Conference on Artificial Intelligence in Education}, pp.~311--322, Springer,
	2017.
	
	\bibitem{Nguyen-PHG:2014}
	A.~Nguyen, C.~Piech, J.~Huang, and L.~Guibas, ``Codewebs: scalable homework
	search for massive open online programming courses,'' in {\em Proceedings of
		the 23rd international conference on World wide web}, pp.~491--502, ACM,
	2014.
	
	\bibitem{Jeuring-BGH:2014}
	J.~Jeuring, L.~T. van Binsbergen, A.~Gerdes, and B.~Heeren, ``Model solutions
	and properties for diagnosing student programs in ask-elle,'' in {\em
		Proceedings of the Computer Science Education Research Conference},
	pp.~31--40, ACM, 2014.
	
	\bibitem{M:2012}
	A.~Mitrovic, ``Fifteen years of constraint-based tutors: what we have achieved
	and where we are going,'' {\em User modeling and user-adapted interaction},
	vol.~22, no.~1-2, pp.~39--72, 2012.
	
	\bibitem{Paassen-JH:2016}
	B.~Paa{\ss}en, J.~Jensen, and B.~Hammer, ``Execution traces as a powerful data
	representation for intelligent tutoring systems for programming.,'' in {\em
		Proceedings of the 9th International Conference on Educational Data Mining},
	pp.~183--190, International Educational Data Mining Society, 2016.
	
	\bibitem{Gulwani-RZ:2014}
	S.~Gulwani, I.~Radi{\v{c}}ek, and F.~Zuleger, ``Feedback generation for
	performance problems in introductory programming assignments,'' in {\em
		Proceedings of the 22nd ACM SIGSOFT International Symposium on Foundations of
		Software Engineering}, pp.~41--51, ACM, 2014.
	
	\bibitem{Chen-NSN:2017}
	Z.~Chen, A.~Nguyen, A.~Schlender, and J.~Ngiam, ``Real-time programming
	exercise feedback in moocs.,'' in {\em Proceedings of the 10th International
		Conference on Educational Data Mining}, pp.~414--415, International
	Educational Data Mining Society, 2017.
	
	\bibitem{Piech-HNPSG:2015}
	C.~Piech, J.~Huang, A.~Nguyen, M.~Phulsuksombati, M.~Sahami, and L.~Guibas,
	``Learning program embeddings to propagate feedback on student code,'' {\em
		arXiv preprint arXiv:1505.05969}, 2015.
	
	\bibitem{Bhatia-S:2016}
	S.~Bhatia and R.~Singh, ``Automated correction for syntax errors in programming
	assignments using recurrent neural networks,'' {\em arXiv preprint
		arXiv:1603.06129}, 2016.
	
	\bibitem{Gupta-KS:2018}
	R.~Gupta, A.~Kanade, and S.~Shevade, ``Deep reinforcement learning for
	programming language correction,'' {\em arXiv preprint arXiv:1801.10467},
	2018.
	
	\bibitem{Movzina-LB:2018}
	M.~Mo{\v{z}}ina, T.~Lazar, and I.~Bratko, ``Identifying typical approaches and
	errors in prolog programming with argument-based machine learning,'' {\em
		Expert Systems with Applications}, vol.~112, pp.~110--124, 2018.
	
	\bibitem{Perelman-GG:2014}
	D.~Perelman, S.~Gulwani, and D.~Grossman, ``Test-driven synthesis for automated
	feedback for introductory computer science assignments,'' {\em Proceedings of
		Data Mining for Educational Assessment and Feedback (ASSESS 2014)}, 2014.
	
	\bibitem{Suzuki-SGHDH:2017}
	R.~Suzuki, G.~Soares, E.~Glassman, A.~Head, L.~D'Antoni, and B.~Hartmann,
	``Exploring the design space of automatically synthesized hints for
	introductory programming assignments,'' in {\em Proceedings of the 2017 CHI
		Conference Extended Abstracts on Human Factors in Computing Systems},
	pp.~2951--2958, ACM, 2017.
	
	\bibitem{Mcbroom-YKC:2018}
	J.~McBroom, K.~Yacef, I.~Koprinska, and J.~R. Curran, ``A data-driven method
	for helping teachers improve feedback in computer programming automated
	tutors,'' in {\em International Conference on Artificial Intelligence in
		Education}, pp.~324--337, Springer, 2018.
	
	\bibitem{Price-ZDLB:2018}
	T.~W. Price, R.~Zhi, Y.~Dong, N.~Lytle, and T.~Barnes, ``The impact of data
	quantity and source on the quality of data-driven hints for programming,'' in
	{\em International Conference on Artificial Intelligence in Education},
	pp.~476--490, Springer, 2018.
	
\end{thebibliography}
\end{document}